\documentclass[aps,pre,amsmath,amssymb]{revtex4-2}

\usepackage{polski}
\usepackage[english]{babel}
\usepackage{bm}
\usepackage{graphicx}
\usepackage{amsmath}
\usepackage{mathrsfs}

\begin{document}

\newcommand{\uu}[1]{\underline{#1}}
\newcommand{\pp}[1]{\phantom{#1}}
\newcommand{\be}{\begin{eqnarray}}
\newcommand{\ee}{\end{eqnarray}}
\newcommand{\ve}{\varepsilon}
\newcommand{\vp}{\varphi}
\newcommand{\vs}{\varsigma}
\newcommand{\Tr}{{\,\rm Tr\,}}
\newcommand{\Trr}{{\,\rm Tr}}
\newcommand{\pol}{\frac{1}{2}}
\newcommand{\sgn}{{\rm sgn}}
\newcommand{\Mo}{\mho}
\newcommand{\Om}{\Omega}

\title{Three-space as a quantum  hyper-layer in 1+3 dimensions}
\author{Marek Czachor}
\affiliation{
Instytut Fizyki i Informatyki Stosowanej,
Politechnika Gdańska, 80-233 Gdańsk, Poland
}

\begin{abstract}
We discuss a formalism where a universe is identified with the support of a wave function propagating through space-time. As opposed to classical cosmology, the resulting universe is not a spacelike section of some space-time, but a hyper-layer of a finite timelike width, a set which is not a three-dimensional submanifold of space-time. We test the formalism on the example of a universe that contains a single harmonic oscillator (a generalization of the curvature-dependent Cari\~nena-Ra\~nada-Santander (CRS) model). As opposed to the original CRS formulation, here the curvature is not a parameter but a quantum observable, a function of the world-position operator. It is shown that asymptotically, for large values of the invariant evolution parameter $\tau$, one reconstructs the standard quantum results, with one modifiication: The effective (renormalized) mass of the oscillator decreases with $\tau$. The effect does not seem to be a peculiarity of harmonic oscillators, so one may speculate that masses of distant elementary quantum systems are greater from their values known from our quantum mechanical measurements.
\end{abstract}
\maketitle

\section{Introduction}

Our human brains have no difficulty imagining a two-dimensional surface, although  ``surfaces'' known from physical experience are objects with non-zero thickness, hence layers, not surfaces. Realistic layers consist of atoms, but quantum mechanics describes atoms as objects that do not possess concrete positions in space. Rather, atoms are in superpositions of different localizations and thus are fundamentally delocalized. In effect, at the most fundamental level, we are always dealing with ``quantum surfaces'' that exist in superpositions of different geometries.

A similar situation is found in space-time physics. What we regard as ``space'' is modeled as a three-dimensional hyper-surface of space-time, an object with zero-thickness in timelike directions.  Yet, our experience of time is fleeting and ephemeral. It is very difficult, if not impossible, to be truly here and  now. In this sense, we do not have everyday experience with space as a three-dimensional submanifold of space-time. The experience of ``now'' seems as delocalized as the atoms that form a quantum surface. Perhaps, what we regard as space is not a hyper-surface, but a hyper-layer. 

This is the first intuition behind the present paper. 
The second intuition is related to the passage of time. 

The passage of time means that even if we accept that ``now'' is somewhat uncertain, the past should eventually disappear and a sufficiently distant future should not yet exist.  This type of behavior does not appear to have an equivalent in standard relativistic physics, but is well known from quantum mechanics. Indeed, a propagating wave packet represents a particle in superposition of different localizations, concentrated around the point of maximal probability density. By Ehrenfest theorem, average position of the wave packet propagates along a solution of a classical Newton equation. A probability of finding the particle far behind, or far ahead of the wave-packet is negligible. Now, it is enough to replace the space coordinate by $x^0=ct$, and treat the evolution parameter, $\tau$ say, as something distinct from $x^0$. A suitable wave packet propagates along $x^0$, and the result is precisely the dynamics where the past literally disappears, the future has not yet happened, and the uncertainty of ``now'' is represented by the width of the packet. A three-dimensional classical hyper-surface is then obtained by an appropriate Ehrenfest theorem, in exactly the same way a classical solution of Newton's equation if found if we follow the average of the position operator.

The idea we have just outlined is not entirely new. It seems that its first explicit formulation was given, for a 1+1 dimensional toy-model,  in \cite{MCAP}. A generalization valid for any dimension was completed in \cite{MC2023} for the case of an empty universe. An inclusion of matter was briefly discussed in \cite{MC2023} as well, but a concrete study of a quantum mechanical system that exists and evolves in such a quantum spacetime is still missing. The present paper is the first attempt of formulating and exactly solving a non-toy model in 1+3. 

For obvious reasons, a harmonic oscillator is our first target. Standard non-relativistic oscillator is simple, well understood, exactly solvable by the factorization method, and is a cornerstone of field quantization. On the other hand, when it comes to relativistic physics, numerous possibilities occur. A generalization based on the Dirac equation was introduced by Moshinsky  and Szczepaniak \cite{Moshinsky1989,Moshinsky}. 
An alternative Dirac-type relativistic harmonic oscillator implicitly occurs in the so-called second Dirac equation \cite{Dirac}, whose deep and exhaustive discussion can be found in \cite{MDB}. A generalization where the kinetic term of the Hamiltonian is proportional to the d'Alembertian $\partial^\mu\partial_\mu$ in Minkowski space, while the potential is proportional to $x_\mu x^\mu$, is discussed in \cite{Horwitz}, generalizing the earlier works of Stueckelberg, Horwitz, and Piron \cite{S1,S2,HP}.  The Horwitz Hamiltonian  is also basically equivalent to Born's ``metric operator'' $p_\mu p^\mu+x_\mu x^\mu$ (in dimensionless units), generating a spectrum of meson masses \cite{Born}. The formalism from \cite{Horwitz} is similar to ours in that the evolution parameter is unrelated to $x^0$. In both formulations, the Hilbert spaces consist of functions integrable with respect to $d^4x$. However, both theories employ different Hamiltonians, different boundary conditions, and different correspondence principles with standard quantum mechanics. Our treatment of the harmonic oscillator is closest to the approach of Cari\~nena, Ra\~nada, and Santander (CRS)  \cite{Carinena2011,Carinena2012}, but the overall structure of our Schr\"odinger equation differs in the form of the free Hamiltonian, with empty universe as a reservoir for matter fields, a missing element of the formalism from \cite{Carinena2011,Carinena2012}. 

We begin in Sec.~\ref{Sec1} with a summary of the formalism proposed in \cite{MC2023}. We concentrate on the distinction between the universe and its background space-time, and on the meaning of the boundary condition. In Sec.~\ref{Sec2} we discuss in some detail  the dynamics of the empty universe. An explicit example, adapted from \cite{MC2023}, illustrates the evolution in 1+1 dimensional background. We also introduce the notion of a configuration-space universe. In Sec.~\ref{Sec3} we give a concrete example of a single oscillator coupled to the universe whose background Minkowski space is 1+3 dimensional. We discuss both Schr\"odinger and interaction pictures. The interaction picture eliminates the free evolution of the universe for the price of making $\tau$-dependent the Hamiltonian of the oscillator. This $\tau$-dependence will manifest itself in the form of the ground state, making the impression that the mass of the oscillator decreases with $\tau$. The effective $\tau$-dependence of mass can be ignored at time scales available in typical quantum measurements, but in principle could influence interpretation of data from very distant objects. The next two sections discuss in detail the ground state of the oscillator. We first analyze in Sec.~\ref{Sec4} an approximation which is simpler to analyze than the exact model from \cite{Carinena2011,Carinena2012}. We concentrate on the ground state, but all the excited states can be found in Appendix B.  We show that the ground state is a Gaussian, but with respect to the geodesic position operator $\hat{\mathtt{r}}=\hat{\mathtt{x}}\otimes\hat\xi$, whose eigenvalues represent the geodesic distance ${\mathtt{r}}={\mathtt{x}}\xi$ computed along the hyperboloid $\mathtt{x}^2=g_{\mu\nu}x^\mu x^\nu$.  In Sec.~\ref{Sec5} we perform an analogous analysis of the ground state for the CRS oscillator. As opposed to the original CRS formulation, the solution we find describes an oscillator in superposition of different and $\tau$-dependent curvatures  of the universe. The curvature that occurs in the solution is not, as opposed to  \cite{Carinena2011,Carinena2012}, a parameter, but a quantum observable, a fact proving that we indeed discuss a quantum oscillator  in a quantum universe. It is also shown that differences between the exact CRS model and its much simpler approximate form are visible only for small $\tau$, that is, in early stages of evolution of our quantum universe. For $\tau$ corresponding to the Hubble time, both models are indistinguishable, a result useful from the point of view of the correspondence principle with standard quantum mechanics.  Sec.~\ref{Sec6} is devoted to the reduction $1+3\to 3$, obtained by integrating out the width of the layer. We show that the resulting probability density is a bell-shaped curve similar to a Gaussian, but more smeared out.  Sec.~\ref{2-body} looks at the above issues from the perspective of a general two-body problem.  Finally, in Appendix A we discuss an alternative definition of the harmonic oscillator associated with a given Laplace-Beltrami operator. Unfortunately, the resulting potential does not lead to any known factorization of the Hamiltonian.

\section{Expanding quantum universe}
\label{Sec1}

In the formalism proposed in \cite{MC2023} a universe is identified with a $\tau$-dependent subset of the background Minkowski space $\cal M$ of  signature $(+-\ldots -)$  in $D$ dimensions. A point $x\in\cal M$ is said to belong to the universe if $\Psi_\tau(x)\neq 0$, where $\Psi_\tau$ is a solution of a certain Schr\"odinger-type equation,  
\be
i\dot \Psi_\tau={\cal H}\Psi_\tau,
\quad
\Psi_\tau=e^{-i{\cal H}\tau }\Psi_{0}.\label{0}
\ee
The evolution parameter $\tau$ is dimensionless. $\Psi_\tau(x)$ is normalized in a $\tau$-invariant way,
\be
\langle \Psi_\tau|\Psi_\tau\rangle
&=&
\int_{V_+}d^Dx |\Psi_\tau(x)|^2=1,\label{1}
\ee
but only for $\tau\ge 0$. For $\tau<0$ the norm can become $\tau$-dependent. Thus, (\ref{1}) simultaneously defines a minimal possible value of $\tau$, and the corresponding arrow of time. Hamiltonian ${\cal H}$ is not self-adjoint, and yet it generates a meaningful unitary dynamics of the universe. The non-self-adjointness is related to the existence of a minimal value of $\tau$. For large $\tau$, the asymptotic evolution of the universe should reconstruct the form of quantum mechanics we know from textbooks. 

$V_+$ consists of future-timelike world-vectors $x^\mu$. The origin of the cone, $x=0$, is arbitrary. In standard classical FRW-type cosmology it would be natural to identify $x=0$ with the Big Bang. In our model, however, the initial condition at $\tau=0$ corresponds to a universe which is not localized at a point in space-time. The initial boundary condition reads $\Psi_{0}(x)=0$ if $x\not\in V_+$. Such a $\Psi_{0}(x)$ vanishes outside of the cone $V_+$, including  $\partial V_+$, the  future light-cone of the origin $x=0$. We additionally assume that the initial wavefunction $\Psi_{0}(x)$ is smooth and vanishes if
$\mathtt{x}=\sqrt{(x^0)^2-(x^1)^2-\ldots-(x^{D-1})^2}$ does not belong to a certain open interval $]A_0,B_0[\subset\mathbb{R}_+$. The possible values of $\tau$ turn out to be related to the choice of $A_0$. Accordingly, different choices of $\Psi_{0}(x)$ imply different limitations on the minimal value of acceptable $\tau$. All these properties would be impossible if $\cal{H}$ were self-adjoint,  so that the non-self-adjointness of the Hamiltonian is an important conceptual ingredient of the theory.

The dynamics introduced in \cite{MC2023} guarantees that the support of $\Psi_{\tau}(x)$ involves only those $x$ whose Minkowskian norm $\mathtt{x}$ belongs to $[A_\tau,B_\tau]\subset\mathbb{R}_+$, with $\lim_{\tau\to\infty}A_\tau=\infty$ and $\lim_{\tau\to\infty}(B_\tau-A_\tau)= 0$ (the support  is the closure of the set where a given function is nonzero, so the closed interval is not a typo). Unitarity of the evolution semigroup thus implies that the universe expands in space and shrinks in time. The unitary dynamics generated by $\cal H$ is of a squeezing type.

The usual spatial section of a FRW-type universe is here replaced by the support of $\Psi_{\tau}(x)$, but as $\tau$ increases the timelike width $B_\tau-A_\tau$ shrinks to 0, so the support gets asymptotically concentrated in a neighborhood of a hyperbolic section in $V_+$. In consequence, the universe is a subset of $\cal M$ which resembles a true material hyperbolic layer of a finite timelike width, propagating towards the future. We assume that at time scales of the order of the Hubble time  the timelike width of the layer is of the order of the Planck length, which leads to the estimate $\tau\sim 10^{243}$ of the current value of $\tau$, whereas one year is of the order of $\tau\sim 10^{203}$ \cite{MC2023}.

An analogous construction can be performed for spherical and flat universes. 

\section{Free dynamics of an empty universe}
\label{Sec2}

The free Hamiltonian that describes an empty universe is given by
\be
{\cal H}_0&=&
-
\frac{\ell^D}{D\mathtt x^D}x^{\mu}\, i\partial_{\mu}.
\label{Omega}
\ee
It generates a unitary semigroup. 
Recall that $\mathtt{x}^2= g_{\mu\nu}x^\mu x^\nu$, where $g_{\mu\nu}$ is the Minkowski-space metric. 
Here $\ell$ is a fundamental length parameter (the Planck length, say). Parametrizing the solution by means of $\mathtt{x}$ and the world-velocity $u^\mu=x^\mu/\mathtt{x}$,
\be
\psi_{\tau}(\mathtt{x},u)
&=&
\Psi_{\tau}(x),
\ee
we find that ${\cal H}_0$ is a generator of translations of the non-negative variable $\mathtt{x}^D$,
\be
{\cal H}_0\psi_{\tau}(\mathtt{x},u)&=&
-i
\ell^D\frac{\partial}{\partial(\mathtt{x}^D)}\psi_{\tau}(\mathtt{x},u),
\label{Omega'}
\ee
so is not self-adjoint, an advantage in this context, as it turns out.

\begin{figure}
\includegraphics[width=8 cm]{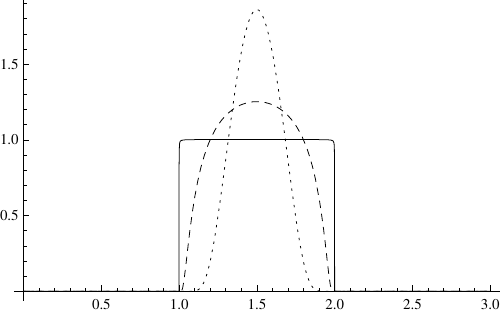}
\caption{Normalized versions of (\ref{10}) for $A_0=1$, $B_0=2$, and $\alpha=1$ (dotted), $\alpha=10$ (dashed), $\alpha=10^4$ (full). With increasing $\alpha$ the function converges pointwise to the characteristic function of the interval $]A_0,B_0[$. For finite $\alpha$ the function is smooth and all its derivatives vanish at both $A_0$ and $B_0$. The plotted functions are normalized with respect to 
$\langle f|f\rangle = \int_0^\infty d\mathtt{x}\,|f(\mathtt{x})|^2$.}
\label{Fig0}
\end{figure}

We assume that $\Psi_{0}(x)$ is smooth and compact-support  in the variable $\mathtt{x}$,  vanishing outside of $]A_0,B_0[\,\subset\mathbb{R}_+$. An example of such a function is 
\be
\exp\frac{1}{\alpha(\mathtt{x}-A_0)(\mathtt{x}-B_0)}, \quad \textrm{for $A_0<\mathtt{x}<B_0$,}\label{10}
\ee
and 0 otherwise. Normalized versions of (\ref{10}) for $A_0=1$, $B_0=2$, and various values of $\alpha$, are depicted in Fig.~\ref{Fig0}. With $\alpha\to\infty$ (\ref{10}) converges pointwise to the characteristic function of $]A_0,B_0[$. In many examples we will tacitly assume that $\alpha$ is so large that (\ref{10}) is practically indistinguishable from the characteristic function, and yet it remains smooth on the whole of $\mathbb{R}_+$.

\begin{figure}
\includegraphics[width=8 cm]{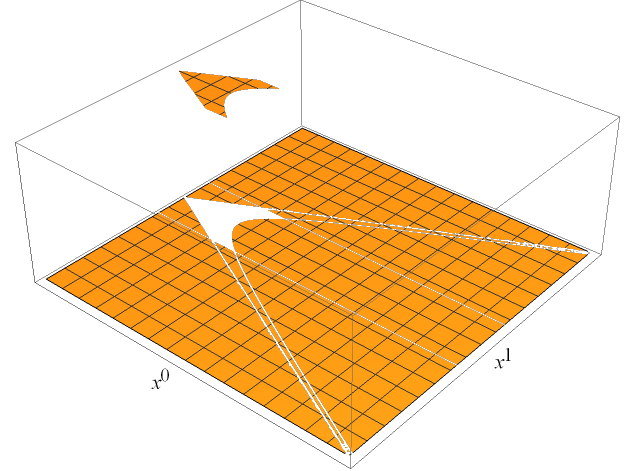}
\includegraphics[width=8 cm]{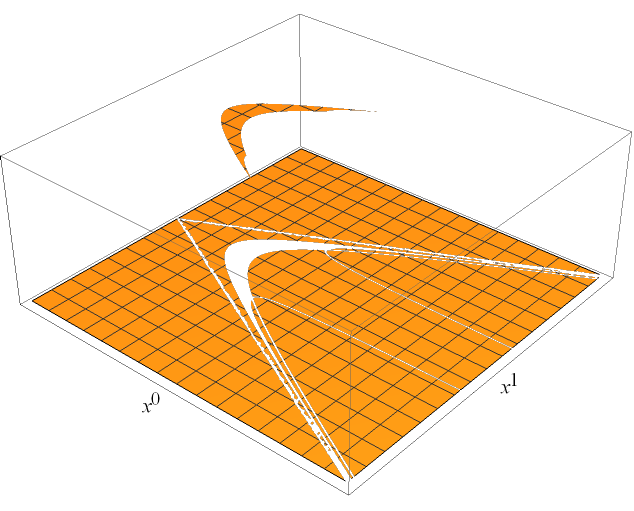}
\caption{Space as an evolving quantum  hyper-layer of space-time \cite{MC2023}. Plot of (\ref{6})--(\ref{7}) with the initial condition (\ref{9}) at $\tau=0$ (left), and its evolved version for $\tau=1$ (right) in a $D=1+1$ Minkowski space, in units where $\ell=1$. $\Psi_\tau(x)$ at $\tau=1$ is thinner and wider than $\Psi_\tau(x)$ at $\tau=0$. A space-time gap occurs between the support of $\Psi_1(x)$ and the light cone. In $1+3$ dimensions the hyperboloid that determines the gap is given by $\ell^2\sqrt{\tau}=c^2t^2-\bm x^2$, as contrasted with 
$c^2t^2-\bm x^2\sim \tau^2$ one might expect on the basis of intuitions developed in special relativity.  Note that the support of $\Psi_\tau(x)$ is not restricted to single hyperbolic space. Rather, $\Psi_\tau(x)$ is a superposition of functions defined on hyperbolic spaces of different curvatures, corresponding to different values of $\mathtt{x}^2=g_{\mu\nu}x^\mu x^\nu$. What we regard as the universe, is the set of those $x^\mu$ in Minkowski space where $\Psi_\tau(x)\neq 0$. The universe is not a $(D-1)$-dimensional submanifold of the Minkowski space, but a $D$-dimensional open subset of the Minkowski space. This is the new paradigm. Harmonic oscillators discussed in the present paper involve less trivial forms of $\Psi_\tau(x)$.}
\label{Fig1}
\end{figure}

Under the above form of the initial condition one finds, for $\ell^D\tau <\mathtt x^D$, 
\be
\Psi_{\tau}(x)
&=&
\Psi_{0}\left(
\left(\frac{\mathtt x^D-\ell^D\tau }{\mathtt x^D}\right)^{1/D}x
\right)\label{6}\\
&=&
e^{-i\tau {\cal H}_0}
\Psi_{0}(x),\label{7}
\ee
whereas for $0\le \mathtt x^D\le \ell^D\tau $ the solution vanishes identically,
\be
\Psi_{\tau}(x)
&\equiv&
0.\label{8}
\ee
The limiting value $\mathtt{x}^D=\ell^D\tau$  defines the {\it gap hyperboloid\/},
\be
\mathtt{x}^2=(x^0)^2-(x^1)^2-\ldots-(x^{D-1})^2=\ell^2\tau^{2/D}.\label{gap}
\ee
An example of such a dynamics in $D=1+1$ is illustrated in Fig.~\ref{Fig1} (adapted from  \cite{MC2023}), with the initial condition
\be
\Psi_0(x)
=
\left\{
\begin{array}{cl}
1 & \textrm{for $|x^1|<1$,  $(x^0)^2-(x^1)^2<1$ , $ x_0>0$}\\
0& \textrm{otherwise} 
\end{array}
\right.
\label{9}
\ee
We assume that the jumps in (\ref{9}) approximate a function of the form (\ref{10}).


Before one fills the universe with matter, one needs an analogous formulation of configuration spaces. 
An $N$-point configuration-space empty-universe extension is defined by
\be
\Psi_{\tau}(x_1,\dots,x_N)
&=&
\Psi_{0}\left(
\left(\frac{\mathtt x_1^D-\ell^D\tau }{\mathtt x_1^D}\right)^{1/D}x_1,\dots,
\left(\frac{\mathtt x_N^D-\ell^D\tau }{\mathtt x_N^D}\right)^{1/D}x_N,
\right)\label{17},
\ee
which is equivalent to
\be
\psi_\tau\left(
\mathtt{x}_1,u_1,\dots,
\mathtt{x}_N,u_N
\right)
&=&
\psi_{0}\left(
\sqrt[D]{\mathtt x_1^D-\ell^D\tau },u_1,\dots,
\sqrt[D]{\mathtt x_N^D-\ell^D\tau },u_N
\right)\label{18}
\ee
with the Hamiltonian
\be
{\cal H}_0&=&
-
\sum_{j=1}^N
\frac{\ell^D}{D\mathtt x_j^D}x_j^{\mu}\, i\frac{\partial}{\partial x_j^{\mu}}
=
-i\ell^D
\sum_{j=1}^N
\frac{\partial}{\partial (\mathtt{x}_j^D)}\label{Omega_N},
\ee
and the norm
\be
\langle \Psi_\tau|\Psi_\tau\rangle
&=&
\int_{V_+}d^Dx_1\dots \int_{V_+}d^Dx_N |\Psi_\tau(x_1,\dots,x_N)|^2\\
&=&
\int_0^\infty d\mathtt{x}_1\,\mathtt{x}_1^{D-1}
\int_{u_1^2=1}du_1
\dots
\int_0^\infty d\mathtt{x}_N\,\mathtt{x}_N^{D-1}
\int_{u_N^2=1}du_N
\,
|\psi_\tau\left(
\mathtt{x}_1,u_1,\dots,
\mathtt{x}_N,u_N
\right)|^2.\label{20}
\ee
Here, $du_j$ denotes an invariant measure on the world-velocity hyperboloid $u_j^2=g_{\mu\nu}u_j^\mu u_j^\nu=1$. 
The right-hand-side of (\ref{Omega_N}) is applicable to functions that involve the world-velocity parametrization of the form (\ref{18}). 
This point is somewhat tricky, as illustrated  in Appendix B by Hamiltonians (\ref{182'''}) and (\ref{183'''}).

Now that we know how to describe the dynamics of an empty universe, it remains to add matter. We will have two goals in mind. First of all, 
we have to test on some well understood quantum mechanical models the asymptotic correspondence principle. The latter means that for very large 
$\tau$ the theory should look like an ordinary quantum mechanics, where the integrals are over $\mathbb{R}^{D-1}$ and not $\mathbb{R}^D$. 
Secondly, 
we should understand if and how the presence of matter influences a geometry of the universe, that is the probability density 
$|\Psi_\tau(x_1,\dots,x_N)|^2$. A nontrivial modification of the Hamiltonian, 
\be
{\cal H}={\cal H}_0\mapsto {\cal H}_0+{\cal H}_1,\label{22}
\ee
can, in principle,  influence the set of points that satisfy $|\Psi_\tau(x_1,\dots,x_N)|^2\neq 0$.
As the universe is the set of those $x$ that have non-zero probability density, the same remark applies to the configuration-space universe.

\section{Quantum oscillator in $D=1+3$}
\label{Sec3}

We skip the intermediate stage of a general two-body problem (postponed till Sec.~\ref{2-body}), and directly concentrate on the configuration space of a relative coordinate. 
To this end, let us parametrize a future-pointing timelike world-vector in (1+3)-dimensional configuration Minkowski space by the ``polar relative coordinates'',
\be
x^0 &=& ct=\mathtt{x}\cosh\xi,\quad \mathtt{x}>0,\,\xi\ge 0,\\
x^1 &=& \mathtt{x}\sinh\xi\sin\theta\cos\varphi,\quad 0\le\theta\le\pi,\, 0\le\varphi<2\pi,\\
x^2 &=& \mathtt{x}\sinh\xi\sin\theta\sin\varphi,\\
x^3 &=& \mathtt{x}\sinh\xi\cos\theta.
\ee
In $D=1+3$, the Hamiltonian consists of the empty-universe free part, ${\cal H}_0$,
\be
{\cal H}_0\psi_{\tau}(\mathtt{x},\xi,\theta,\varphi)&=&
-i
\ell^4\frac{\partial}{\partial(\mathtt{x}^4)}\psi_{\tau}(\mathtt{x},\xi,\theta,\varphi),
\label{23}
\ee
and the matter-field interaction part,
\be
{\cal H}_1 &=& \epsilon H,
\ee
where $H$ is some ``ordinary'' Hamiltonian that describes a quantum system on the hyperboloid $\mathtt{x}^2=g_{\mu\nu}x^\mu x^\nu>0$. $\epsilon$ is a parameter that makes ${\cal H}_1$ as dimensionless as  $\tau$.  We shall concentrate on $H$  describing some form of a relativistic harmonic oscillator.

The Minkowski-space metric satisfies
\be
g_{\mu\nu}dx^\mu dx^\nu
&=&
(dx^0)^2-(dx^1)^2-(dx^2)^2-(dx^3)^2\\
&=&
d\mathtt{x}^2
-
\mathtt{x}^2d\xi^2
-
\mathtt{x}^2
\sinh^2\xi \,d\theta^2
-
\mathtt{x}^2
\sinh^2\xi \sin^2\theta \,d\varphi^2.
\ee
The corresponding Jacobian, 
\be
d^4x
&=&
dx^0dx^1dx^2dx^3\\
&=&
d\mathtt{x}\,d\xi \,d\theta \,d\varphi\,\mathtt{x}^3\sinh^2\xi\sin\theta,
\ee
is consistent with Hermiticity of the polar-form d'Alembertian,
\be
\Box &=& 
\frac{\partial^2}{\partial (x^0)^2}
-
\frac{\partial^2}{\partial (x^1)^2}
-
\frac{\partial^2}{\partial (x^2)^2}
-
\frac{\partial^2}{\partial (x^3)^2}
\\
&=&
\frac{1}{\mathtt{x}^3}\frac{\partial}{\partial\mathtt{x}}\mathtt{x}^3\frac{\partial}{\partial\mathtt{x}}
-
\frac{1}{\mathtt{x}^2\sinh^2\xi}\frac{\partial}{\partial\xi}\sinh^2\xi\frac{\partial}{\partial\xi}
-
\frac{1}{\mathtt{x}^2\sinh^2\xi}
\left(
\frac{1}{\sin\theta}\frac{\partial}{\partial\theta}\sin\theta\frac{\partial}{\partial\theta}
+
\frac{1}{\sin^2\theta}\frac{\partial^2}{\partial\varphi^2}
\right),
\ee
if appropriate boundary conditions are imposed.
The operator,
\be
\Delta_\mathtt{x}
&=&
\frac{1}{\mathtt{x}^2\sinh^2\xi}\frac{\partial}{\partial\xi}\sinh^2\xi\frac{\partial}{\partial\xi}
+
\frac{1}{\mathtt{x}^2\sinh^2\xi}
\left(
\frac{1}{\sin\theta}\frac{\partial}{\partial\theta}\sin\theta\frac{\partial}{\partial\theta}
+
\frac{1}{\sin^2\theta}\frac{\partial^2}{\partial\varphi^2}
\right),\label{32'}
\ee
is the Laplace-Beltrami operator on the hyperboloid $\mathtt{x}^2=g_{\mu\nu}x^\mu x^\nu>0$.

There are numerous ways of defining a harmonic oscillator in hyperbolic geometry (cf. \cite{Moshinsky,Horwitz} and Appendix A), but  we find it simplest to consider a potential proportional to $\tanh^2\xi$ \cite{Carinena2011,Carinena2012},
\be
H
&=&
-\frac{\hbar^2}{2\mu}
\Delta_\mathtt{x}
+
\frac{\mu\omega^2\mathtt{x}^2\tanh^2\xi}{2}.
\ee
A world-vector $(ct,\bm x)$ satisfies in polar coordinates $r^2=\bm x^2=\mathtt{x}^2\sinh^2\xi$, so the potential 
\be
\frac{\mu\omega^2\mathtt{x}^2\sinh^2\xi}{2\cosh^2\xi}
&=&
\frac{\mu\omega^2r^2}{2}\frac{1}{1+\sinh^2\xi}
=
\frac{\mu\omega^2r^2}{2}\frac{1}{1+r^2/\mathtt{x}^2}
\approx
\frac{\mu\omega^2r^2}{2},
\ee
reconstructs the usual harmonic oscillator 
if $r$ is of the size available in present-day experiments, while $\mathtt{x}$ is of the order of the Hubble radius of the universe. This type of approximation agrees with the one for the measure on the hyperboloid, 
\be
\frac{d^3x}{\sqrt{1+\bm x^2/\mathtt{x}^2}}
\approx d^3x,
\ee
which characterizes the correspondence principle with standard quantum mechanics. 

For small values of $\xi$, the potential takes another interesting form, namely
\be
\frac{\mu\omega^2\mathtt{x}^2\tanh^2\xi}{2}
\approx
\frac{\mu\omega^2\mathtt{x}^2\xi^2}{2}
=
\frac{\mu\omega^2\mathtt{r}^2}{2},\label{38'''}
\ee
where $\mathtt{r}=\mathtt{x}\xi$ is the geodesic distance computed along the hyperboloid. Actually, the right-hand side of (\ref{38'''}) is a natural alternative definition of the potential if one assumes that the physical distance between interacting objects should be given in terms of the geodesic distance $\mathtt{r}$, and not in terms of $r=|\bm x|$, as the latter is not a geometrically intrinsic characteristic of the hyperboloid. 

Perhaps we can get a more illuminating picture of the potential by writing it as follows:
\be
\frac{\mu\omega^2\mathtt{x}^2\sinh^2\xi}{2\cosh^2\xi}
&=&
\frac{\mu\omega^2\mathtt{x}^2}{2}\frac{(x^1)^2+(x^2)^2+(x^3)^2}{(x^0)^2},
\ee
with $(x^0)^2>(x^1)^2+(x^2)^2+(x^3)^2$, showing that the possible three-space position of the oscillator is limited by the light cone $\mathtt{x}=0$, the boundary of the background space-time. 

The full Schr\"odinger equation,
\be
i\dot\psi_{\tau}(\mathtt{x},\xi,\theta,\varphi)
&=&
-i
\ell^4\frac{\partial}{\partial(\mathtt{x}^4)}\psi_{\tau}(\mathtt{x},\xi,\theta,\varphi)
+
\epsilon
\left(
-\frac{\hbar^2}{2\mu}
\Delta_\mathtt{x}
+
\frac{\mu\omega^2\mathtt{x}^2\tanh^2\xi}{2}
\right)
\psi_{\tau}(\mathtt{x},\xi,\theta,\varphi),
\ee
can be partly separated by means of 
\be
\psi_{lm,\tau}(\mathtt{x},\xi,\theta,\varphi)
&=&
\phi_{l,\tau}(\mathtt{x},\xi)Y_{lm}(\theta,\varphi),
\ee
\be
i\dot\phi_{l,\tau}(\mathtt{x},\xi)
&=&
-i
\ell^4\frac{\partial}{\partial(\mathtt{x}^4)}\phi_{l,\tau}(\mathtt{x},\xi)
\nonumber\\
&\pp=&
+
\epsilon
\left(
-\frac{\hbar^2}{2\mu}
\frac{1}{\mathtt{x}^2\sinh^2\xi}\frac{\partial}{\partial\xi}\sinh^2\xi\frac{\partial}{\partial\xi}
+
\frac{\hbar^2}{2\mu}
\frac{l(l+1)}{\mathtt{x}^2\sinh^2\xi}
+
\frac{\mu\omega^2\mathtt{x}^2\tanh^2\xi}{2}
\right)
\phi_{l,\tau}(\mathtt{x},\xi),
\ee
because the angular momentum operator,
\be
\bm J^2 =
-\hbar^2
\left(
\frac{1}{\sin\theta}\frac{\partial}{\partial\theta}\sin\theta\frac{\partial}{\partial\theta}
+
\frac{1}{\sin^2\theta}\frac{\partial^2}{\partial\varphi^2}
\right),
\ee
commutes with the total Hamiltonian, ${\cal H}={\cal H}_0+{\cal H}_1$.

The free Hamiltonian, ${\cal H}_0$, is a generator of translations of the non-negative variable $\mathtt{x}^4$. It effectively replaces $\mathtt{x}$ by $\mathtt{x}(\tau)=\sqrt[4]{\mathtt{x}^4+ \ell^4\tau}$. The next step is therefore the transition to the interaction picture,
\be
\Phi_{l,\tau}(\mathtt{x},\xi)
&=&
e^{\tau\ell^4\frac{\partial}{\partial(\mathtt{x}^4)}}
\phi_{l,\tau}(\mathtt{x},\xi)
=
\phi_{l,\tau}(\mathtt{x}(\tau),\xi).
\ee
The equation to solve,
\be
i\dot\Phi_{l,\tau}(\mathtt{x},\xi)
&=&
\epsilon
\left(
-\frac{\hbar^2}{2\mu}
\frac{1}{\mathtt{x}(\tau)^2\sinh^2\xi}\frac{\partial}{\partial\xi}\sinh^2\xi\frac{\partial}{\partial\xi}
+
\frac{\hbar^2}{2\mu}
\frac{l(l+1)}{\mathtt{x}(\tau)^2\sinh^2\xi}
+
\frac{\mu\omega^2\mathtt{x}(\tau)^2\tanh^2\xi}{2}
\right)
\Phi_{l,\tau}(\mathtt{x},\xi),\label{46'''}
\ee
is equivalent to a harmonic oscillator on a space of constant but $\tau$-dependent negative curvature. Recall that the solution is normalized by means of
\be
\langle \Phi|\Phi\rangle
&=&
\int_0^\infty d\mathtt{x}\,\mathtt{x}^3\int_0^\infty d\xi \,\sinh^2\xi\,
|\Phi(\mathtt{x},\xi)|^2.\label{47}
\ee
Notice that not only is the curvature $\tau$-dependent, but  it is not a classical parameter, as opposed to the standard literature of the subject. This is a quantum observable, as quantum as the position operator, since one integrates over $\mathtt{x}$ in (\ref{47}). This universe is truly quantum and dynamic. It exists in superposition of different curvatures.

\section{Interlude: Ground state for small $\xi$}
\label{Sec4}

Although Schr\"odinger equation (\ref{46'''}) can be solved exactly, let us first concentrate on the approximation valid for small $\xi$, as it will help us to develop physical intuitions concerning the nature of the solution. Setting $\sinh\xi\approx\tanh\xi\approx\xi$ and $l=0$ (as we search for the ground state) we obtain
\be
i\dot\Phi_{0,\tau}(\mathtt{x},\xi)
&=&
\epsilon
\left(
-\frac{\hbar^2}{2\mu}
\frac{1}{\mathtt{x}(\tau)^2\xi^2}\frac{\partial}{\partial\xi}\xi^2\frac{\partial}{\partial\xi}
+
\frac{\mu\omega^2\mathtt{x}(\tau)^2\xi^2}{2}
\right)
\Phi_{0,\tau}(\mathtt{x},\xi),\label{48}
\ee
with the normalization 
\be
\langle \Phi|\Phi\rangle
&=&
\int_0^\infty d\mathtt{x}\,\mathtt{x}^3\int_0^\infty d\xi \,\xi^2\,
|\Phi(\mathtt{x},\xi)|^2=1.\label{49a}
\ee
The form (\ref{49a}) of the scalar product had to be modified in order to maintain the Hermiticity of the Laplacian in (\ref{48}).

Note that  (\ref{48})--(\ref{49a}) can be  alternatively interpreted as an exact model in a spatially flat universe, where spacelike distances are computed by means of the hyperbolic geodesic distances. Such a flat universe is not equivalent to the Minkowski space, and yet employs the Minkowski space as its background space-time --- an interesting option to contemplate in some future work, especially in the context of $\Lambda$CDM cosmology. 

Now, define $F_{\tau}(\mathtt{x},\xi)=\Phi_{0,\tau}(\mathtt{x},\xi)\xi$. Then
\be
i\dot F_{\tau}(\mathtt{x},\xi)
&=&
\epsilon
\left(
-\frac{\hbar^2}{2\mu}
\frac{1}{\mathtt{x}(\tau)^2}\frac{\partial^2}{\partial\xi^2}
+
\frac{\mu\omega^2\mathtt{x}(\tau)^2\xi^2}{2}
\right)
F_{\tau}(\mathtt{x},\xi)\label{49}
\\
&=&
\epsilon\hbar\omega
\left(
a(\mathtt{x},\tau)^\dag a(\mathtt{x},\tau)
+
\frac{1}{2}
\right)
F_{\tau}(\mathtt{x},\xi)\label{49}.
\ee
The creation and annihilation operators,
\be
a(\mathtt{x},\tau)F_{\tau}(\mathtt{x},\xi)
&=&
\frac{1}{\sqrt{\hbar\omega}}
\left(
\frac{\hbar}{\sqrt{2\mu}}
\frac{1}{\mathtt{x}(\tau)}\frac{\partial}{\partial\xi}
+
\sqrt{\frac{\mu}{2}}\omega\mathtt{x}(\tau)\xi
\right)F_{\tau}(\mathtt{x},\xi),\\
a(\mathtt{x},\tau)^\dag F_{\tau}(\mathtt{x},\xi)
&=&
\frac{1}{\sqrt{\hbar\omega}}
\left(
-\frac{\hbar}{\sqrt{2\mu}}
\frac{1}{\mathtt{x}(\tau)}\frac{\partial}{\partial\xi}
+
\sqrt{\frac{\mu}{2}}\omega\mathtt{x}(\tau)\xi
\right)F_{\tau}(\mathtt{x},\xi),
\ee
satisfy the usual algebra,
\be
{[a(\hat{\mathtt{x}},\tau),a(\hat{\mathtt{x}},\tau)^\dag]}
&=&
\mathbb{I}.
\ee
The hat in $\hat{\mathtt{x}}$ reminds us that ${\mathtt{x}}$ is an eigenvalue of $\hat{\mathtt{x}}$. 
The ground state satisfies $a(\mathtt{x},\tau)F_{0,\tau}(\mathtt{x},\xi)=0$,
\be
\frac{\partial F_{0,\tau}(\mathtt{x},\xi)}{\partial\xi}
&=&
-\frac{\mu\omega}{\hbar}\xi\sqrt{\mathtt{x}^4+\ell^4\tau}
F_{0,\tau}(\mathtt{x},\xi)
,\\
i\dot F_{0,\tau}(\mathtt{x},\xi)
&=&
\frac{\epsilon\hbar\omega}{2}
F_{0,\tau}(\mathtt{x},\xi),
\ee
and thus
\be
F_{0,\tau}(\mathtt{x},\xi)
&=&
F_{0,\tau}(\mathtt{x},0)
e^{-\frac{\mu\omega}{2\hbar}\xi^2\sqrt{\mathtt{x}^4+\ell^4\tau}},
\\
F_{0,\tau}(\mathtt{x},\xi)
&=&
e^{-i\frac{\omega}{2}\epsilon\hbar\tau}
F_{0,0}(\mathtt{x},\xi)
\\
&=&
e^{-i\frac{\omega}{2}\epsilon\hbar\tau}
F_{0,0}(\mathtt{x},0)
e^{-\frac{\mu\omega}{2\hbar}\mathtt{x}^2\xi^2},
\ee
or, equivalently,
\be
\Phi_{0,\tau}(\mathtt{x},\xi)
&=&
e^{-i\frac{\omega}{2}\epsilon\hbar\tau}
F_{0,0}(\mathtt{x},0)
\xi^{-1}e^{-\frac{\mu\omega}{2\hbar}\mathtt{x}^2\xi^2}
\\
&=&
\phi_{0,\tau}\left(\sqrt[4]{\mathtt{x}^4+\ell^4\tau},\xi\right).
\ee
Returning to the Schr\"odinger picture, we finally find
\be
\phi_{0,\tau}(\mathtt{x},\xi)
&=&
e^{-i\frac{\omega}{2}\epsilon\hbar\tau}
F_{0,0}\left(\sqrt[4]{\mathtt{x}^4-\ell^4\tau},0\right)
\xi^{-1}e^{-\frac{\mu\omega}{2\hbar}\sqrt{\mathtt{x}^4-\ell^4\tau}\xi^2}.
\ee
Let us recall that 
$\phi_{0,0}\left(\mathtt{x},\xi\right)$ is non-zero only if $\mathtt{x}\in\,]A_0,B_0[\,\subset \mathbb{R}_+$, for some $A_0$ and $B_0$, a fact implying that 
\be
\langle \phi_{0,\tau}|\phi_{0,\tau}\rangle
&=&
\int_0^\infty d\mathtt{x}\,\mathtt{x}^3\int_0^\infty d\xi \,\xi^2\,
|\phi_{0,\tau}(\mathtt{x},\xi)|^2
\\
&=&
\frac{1}{4}\int_{A_0^4+\ell^4\tau}^{B_0^4+\ell^4\tau} d(\mathtt{x}^4)\int_0^\infty d\xi \,
\left|F_{0,0}\left(\sqrt[4]{\mathtt{x}^4-\ell^4\tau},0\right)\right|^2
e^{-\frac{\mu\omega}{\hbar}\sqrt{\mathtt{x}^4-\ell^4\tau}\xi^2}
\\
&=&
\int_{A_0}^{B_0} d\mathtt{x}\,\mathtt{x}^3\left|F_{0,0}\left(\mathtt{x},0\right)\right|^2\int_0^\infty d\xi \,
e^{-\frac{\mu\omega}{\hbar}\mathtt{x}^2\xi^2}
\label{65n}
\\
&=&
\frac{1}{2}\sqrt{\frac{\pi\hbar}{\mu\omega}}
\int_{A_0}^{B_0} d\mathtt{x}\,\mathtt{x}^2\left|F_{0,0}\left(\mathtt{x},0\right)\right|^2=1.
\ee
In order to simplify the discussion assume that $F_{0,0}\left(\mathtt{x},0\right)$ is given by a function proportional to (\ref{10}) with a large value of $\alpha$, say $\alpha=10^{100}$, so that $F_{0,0}\left(\mathtt{x},0\right)$, being smooth, is practically indistinguishable from the multiple $C\chi_{]A_0,B_0[}(\mathtt{x})$ of the characteristic function $\chi_{]A_0,B_0[}$ of the open interval $]A_0,B_0[$. The normalization now reads
\be
1&\approx&
\frac{|C|^2}{6}\sqrt{\frac{\pi\hbar}{\mu\omega}}
(B_0^3-A_0^3)
\ee
so that
\be
\phi_{0,\tau}(\mathtt{x},\xi)
&\approx&
\sqrt{\frac{6}{B_0^3-A_0^3}}\left(\frac{\mu  \omega }{\pi\hbar}\right)^{1/4}
e^{-i\frac{\omega}{2}\epsilon\hbar\tau}
\chi_{]A_0,B_0[}\left(\sqrt[4]{\mathtt{x}^4-\ell^4\tau}\right)
\xi^{-1}e^{-\frac{\mu\omega}{2\hbar}\mathtt{x}^2\xi^2\sqrt{1-\frac{\ell^4\tau}{\mathtt{x}^4}}}.\label{73}
\ee
The universe consists here of those events whose probability density is non-zero,  $|\phi_{0,\tau}(\mathtt{x},\xi)|^2\neq 0$. Therefore, when analyzing the differences between (\ref{73})  and the standard quantum prediction for the ground state, we can skip the characteristic function, still keeping in mind that its argument satisfies $A_0<\sqrt[4]{\mathtt{x}^4-\ell^4\tau}<B_0$, with $A_0$ and $B_0$ determined by the initial condition for the universe at $\tau=0$, hence some 13 billion years ago. Moreover, it is clear that the role of non-relativistic  time $\mathtt{t}$ is played here by $\epsilon\hbar\tau$, while the product $\mathtt{x}^2\xi^2=\mathtt{r}^2$ is the square of the hyperboloid's geodesic distance. The characteristic function implies that 
\be
0<\frac{A_0^2}{\mathtt{x}^2}<\sqrt{1-\frac{\ell^4\tau}{\mathtt{x}^4}}<\frac{B_0^2}{\mathtt{x}^2}<\frac{B_0^2}{\ell^2\sqrt{\tau}},\label{74}
\ee
so $\phi_{0,\tau}(\mathtt{x},\xi)$ spreads along spacelike directions in a future-neighborhood of the gap hyperboloid $\mathtt{x}^2=\ell^2\sqrt{\tau}$, simultaneously 
shrinking in the timelike direction in a way determined by (\ref{74}). All these properties are consistent with the analysis given in \cite{MC2023}.

A qualitatively new element is given by the square root occurring in the Gaussian, 
\be
-\frac{\mu\omega}{2\hbar}\mathtt{r}^2\sqrt{1-\frac{\ell^4\tau}{\mathtt{x}^4}},
\ee
because, as a consequence of (\ref{74}),  we  effectively find
\be
\lim_{\tau\to\infty}\frac{\mu\omega}{\hbar}\sqrt{1-\frac{\ell^4\tau}{\mathtt{x}^4}}=0.
\ee
Assuming that $\hbar$ is a fundamental constant, and taking into account that $\omega$ occurs in the oscillating term $e^{-i\frac{\omega}{2}\epsilon\hbar\tau}=e^{-i\frac{\omega}{2}\mathtt{t}}$ in exactly the same way as the one we know from textbook quantum mechanics, we conclude that the Gaussian behavior of the geodesic variable $\mathtt{r}=\mathtt{x}\xi$ is controlled by the mass term
\be
\mu\sqrt{1-\frac{\ell^4\tau}{\mathtt{x}^4}},
\ee
which, accordingly, should be observed as decreasing with time. Obviously, in time scales available in present-day quantum measurements we can assume that 
\be
\mu\sqrt{1-\frac{\ell^4(\tau+\delta\tau)}{\mathtt{x}^4}}\approx \mu\sqrt{1-\frac{\ell^4\tau}{\mathtt{x}^4}},
\ee
and thus quantum oscillators are expected to behave as if their masses were time invariant. However, if what we observe is indeed the geodesic position $\mathtt r$, then very distant objects should behave as if their masses were greater from the ones we know from our human laboratory measurements. Our conclusion is reminiscent of some results on time dependent masses of quantized scalar fields in both classical  \cite{Peter,Uzan} and quantum cosmology \cite{mass0,mass1,mass2}.

All we have written above applies to the geodesic observable $\hat{\mathtt r}=\hat{\mathtt{x}}\otimes \hat\xi$, whose eigenvalues are given by $\mathtt r=\mathtt{x}\xi$. A measurement of $\hat{\mathtt r}$ is therefore a measurement of a tensor product of two observables. One of them, namely $\hat{\mathtt{x}}$, determines location of the hyperboloid in the background Minkowski space (up to the uncertainty relation following from (\ref{74})). This is effectively a {\it measurement of quantum time\/}, as it approximately determines the value of $\tau$. The measurement of $\hat\xi$ determines the position of the oscillator along the hyperboloid, so this is, essentially,  a measurement of position. More precisely, the variable $\xi$ has the status of a shape variable, in Barbour's sense (see \cite{Barbour}, and the example discussed in \cite{MC2023}).

The two observables are distributed in spacetime by means of the two reduced probability densities,
\be
\rho_\tau(\mathtt{x})
&=&
\int_0^\infty d\xi \,\mathtt{x}^3\xi^2\,
|\phi_{0,\tau}(\mathtt{x},\xi)|^2,\quad \textrm{(probability density of quantum time)},\label{79}\\
\rho_\tau(\xi)
&=&
\int_0^\infty d\mathtt{x}\,\mathtt{x}^3\xi^2\,
|\phi_{0,\tau}(\mathtt{x},\xi)|^2=\rho_0(\xi),\quad \textrm{(probability density of quantum position)}.\label{80}
\ee
Neither of them is the usual Gaussian, but the joint space-time probability distribution is Gaussian.

\section{The exact ground state}
\label{Sec5}

Let us consider the exact $l=0$ equation (\ref{46'''}) for $F_{\tau}(\mathtt{x},\xi)=\Phi_{0,\tau}(\mathtt{x},\xi)\sinh\xi$,
\be
i\dot F_{\tau}(\mathtt{x},\xi)
&=&
\epsilon
\frac{\hbar^2}{2\mu}
\frac{1}{\mathtt{x}(\tau)^2}
\left(
-
\frac{\partial^2}{\partial\xi^2}
+
\frac{\mu^2\omega^2\mathtt{x}(\tau)^4\tanh^2\xi}{\hbar^2}
+
1\right)
F_{\tau}(\mathtt{x},\xi).\label{76a}
\ee
The Hamiltonian in (\ref{76a}) can be factorized,
\be
i\dot F_{\tau}(\mathtt{x},\xi)
&=&
\epsilon
\frac{\hbar^2}{2\mu}
\frac{1}{\mathtt{x}(\tau)^2}
\left(
A(\mathtt{x},\tau)^\dag A(\mathtt{x},\tau)
+\sqrt{\frac{1}{4}+\frac{\mu^2\omega^2\mathtt{x}(\tau)^4}{\hbar^2}}
+\frac{1}{2}
\right)
F_{\tau}(\mathtt{x},\xi)
,\\
A(\mathtt{x},\tau)
&=&
\frac{\partial}{\partial\xi}+\left(\sqrt{\frac{1}{4}+\frac{\mu^2\omega^2\mathtt{x}(\tau)^4}{\hbar^2}}-\frac{1}{2}\right) \tanh\xi,\\
A(\mathtt{x},\tau)^\dag
&=&
-\frac{\partial}{\partial\xi}+\left(\sqrt{\frac{1}{4}+\frac{\mu^2\omega^2\mathtt{x}(\tau)^4}{\hbar^2}}-\frac{1}{2}\right) \tanh\xi
.
\ee
The ground state is annihilated by $A(\hat{\mathtt{x}},\tau)$, 
\be
A(\mathtt{x},\tau)F_{0,\tau}(\mathtt{x},\xi),
&=&
0,
\ee
which implies
\be
\frac{\partial F_{0,\tau}(\mathtt{x},\xi)}{\partial\xi}
&=&
-\left(\sqrt{\frac{1}{4}+\frac{\mu^2\omega^2\mathtt{x}(\tau)^4}{\hbar^2}}-\frac{1}{2}\right) \tanh\xi
F_{0,\tau}(\mathtt{x},\xi)
\ee
and
\be
\Phi_{0,\tau}(\mathtt{x},\xi)
&=&
\frac{F_{0,\tau}(\mathtt{x},\xi)}{\sinh\xi}\\
&=&
\frac{1}{\sinh\xi}(\cosh\xi)^{\frac{1}{2}-\sqrt{\frac{1}{4}+\frac{\mu^2\omega^2(\mathtt{x}^4+\ell^4\tau)}{\hbar^2}}}
F_{0,\tau}(\mathtt{x},0).\label{83}
\ee
Let us keep in mind that this is still the interaction-picture solution.  

The $\tau$-dependent equation,
\be
i\dot F_{0,\tau}(\mathtt{x},\xi)
&=&
\epsilon
\frac{\hbar^2}{2\mu}
\frac{1}{\mathtt{x}(\tau)^2}
\left(
\sqrt{\frac{1}{4}+\frac{\mu^2\omega^2\mathtt{x}(\tau)^4}{\hbar^2}}
+\frac{1}{2}
\right)
F_{0,\tau}(\mathtt{x},\xi),
\ee
is solved by
\be
F_{0,\tau}(\mathtt{x},\xi)
&=&
e^{
-i\epsilon
\frac{\hbar^2}{4\mu}
\int_0^\tau d\tau'\left(
\sqrt{\frac{1}{\mathtt{x}^4+\ell^4\tau'}+\frac{4\mu^2\omega^2}{\hbar^2}}
+\frac{1}{\sqrt{\mathtt{x}^4+\ell^4\tau'}}
\right)}
F_{0,0}(\mathtt{x},\xi),
\ee
with $F_{0,0}(\mathtt{x},\xi)$ following from (\ref{83}),
\be
F_{0,0}(\mathtt{x},\xi)
&=&
(\cosh\xi)^{\frac{1}{2}-\sqrt{\frac{1}{4}+\frac{\mu^2\omega^2\mathtt{x}^4}{\hbar^2}}}
F_{0,0}(\mathtt{x},0).\label{86}
\ee
The integral
\be
{\cal I}(\mathtt{x},\omega,\tau)
&=&
\int_0^\tau d\tau'
\sqrt{\frac{1}{\mathtt{x}^4+\ell^4\tau'}+\frac{4\mu^2\omega^2}{\hbar^2}}
\ee
is explicitly given by
\be
{\cal I}(\mathtt{x},\omega,\tau)
&=&
\frac{\hbar}{2\mu\omega}
\frac{1}{\ell^4}
\ln\left(
\sqrt{1+\frac{4\mu^2\omega^2}{\hbar^2}(\mathtt{x}^4+\ell^4\tau)}+\frac{2\mu\omega}{\hbar}\sqrt{\mathtt{x}^4+\ell^4\tau}
\right)
+
\frac{1}{\ell^4}
\sqrt{1+\frac{4\mu^2\omega^2}{\hbar^2}(\mathtt{x}^4+\ell^4\tau)}\sqrt{\mathtt{x}^4+\ell^4\tau}
\nonumber\\
&\pp=&
-
\frac{\hbar}{2\mu\omega}
\frac{1}{\ell^4}
\ln\left(
\sqrt{1+\frac{4\mu^2\omega^2}{\hbar^2}\mathtt{x}^4}+\frac{2\mu\omega}{\hbar}\mathtt{x}^2
\right)
-
\frac{1}{\ell^4}
\sqrt{1+\frac{4\mu^2\omega^2}{\hbar^2}\mathtt{x}^4}\mathtt{x}^2.\label{88}
\ee
Note that 
\be
\lim_{\omega\to 0}
\frac{\hbar}{2\mu\omega}
\frac{1}{\ell^4}
\ln\left(
\sqrt{1+\frac{4\mu^2\omega^2}{\hbar^2}(\mathtt{x}^4+\ell^4\tau)}+\frac{2\mu\omega}{\hbar}\sqrt{\mathtt{x}^4+\ell^4\tau}
\right)
&=&
\lim_{\omega\to 0}
\frac{1}{\ell^4}
\sqrt{1+\frac{4\mu^2\omega^2}{\hbar^2}(\mathtt{x}^4+\ell^4\tau)}\sqrt{\mathtt{x}^4+\ell^4\tau}
\nonumber\\
&=&
\frac{\sqrt{\mathtt{x}^4+\ell^4\tau}}{\ell^4}\label{89},
\ee
hence,
\be
{\cal I}(\mathtt{x},0,\tau)
&=&
\frac{2}{\ell^4}
\left(
\sqrt{\mathtt{x}^4+\ell^4\tau}
-
\mathtt{x}^2
\right)
=
\int_0^\tau d\tau'
\frac{1}{\sqrt{\mathtt{x}^4+\ell^4\tau'}},
\ee
as implied by (\ref{88})--(\ref{89}), can be cross-checked by direct integration. The full interaction-picture solution reads
\be
\Phi_{0,\tau}(\mathtt{x},\xi)
&=&
e^{
-i\epsilon
\frac{\hbar^2}{4\mu}
\big({\cal I}(\mathtt{x},\omega,\tau)+{\cal I}(\mathtt{x},0,\tau\big)}
\frac{1}{\sinh\xi}(\cosh\xi)^{\frac{1}{2}-\sqrt{\frac{1}{4}+\frac{\mu^2\omega^2\mathtt{x}^4}{\hbar^2}}}
F_{0,0}(\mathtt{x},0),
\ee
which translates in the Schr\"odinger picture into
\be
\phi_{0,\tau}(\mathtt{x},\xi)
&=&
e^{
-i\epsilon
\frac{\hbar^2}{4\mu}
\big({\cal I}(\sqrt[4]{\mathtt{x}^4-\ell^4\tau},\omega,\tau)+{\cal I}(\sqrt[4]{\mathtt{x}^4-\ell^4\tau},0,\tau)\big)}
\frac{1}{\sinh\xi}
(\cosh\xi)^{\frac{1}{2}-\frac{1}{2}\sqrt{1+\frac{4\mu^2\omega^2(\mathtt{x}^4-\ell^4\tau)}{\hbar^2}}}
F_{0,0}\big(\sqrt[4]{\mathtt{x}^4-\ell^4\tau},0\big).\label{92'}
\ee
For $\mathtt{x}>0$, which we assume, the solution is normalized by
\be
\langle \phi_{0,\tau}|\phi_{0,\tau}\rangle
&=&
\langle \phi_{0,0}|\phi_{0,0}\rangle=1\\
&=&
\int_0^\infty d\mathtt{x}\,\mathtt{x}^3\int_0^\infty d\xi \,\sinh^2\xi\,
|\phi_{0,\tau}(\mathtt{x},\xi)|^2\\
&=&
\int_{A_0}^{B_0} d\mathtt{x}\,\mathtt{x}^3\,|F_{0,0}(\mathtt{x},0)|^2
\int_0^\infty d\xi \,
(\cosh\xi)^{1-\sqrt{1+\frac{4\mu^2\omega^2\mathtt{x}^4}{\hbar^2}}}
,
\label{92}
\ee
with 
\be
\int_0^\infty d\xi \,
(\cosh\xi)^{1-a}
&=&
2^{a-1} 
\left(
\frac{_2F_1\left(\frac{1}{2}(a +1),a;\frac{1}{2}(a +3);-1\right)}{a+1}
+
\frac{_2F_1\left(\frac{1}{2}(a-1),a;\frac{1}{2}(a+1);-1\right)}{a-1}\right),
\ee
for $a>1$. (\ref{92}) reconstructs the approximate result (\ref{65n}) in consequence of the limit
\be
\lim_{\mathtt{x}\to\infty}
\big(\cosh(\mathtt{r}/\mathtt{x})\big)^{1-\sqrt{1+\frac{4\mu^2\omega^2\mathtt{x}^4}{\hbar^2}}}
&=&
e^{-\frac{\mu\omega}{\hbar}\mathtt{r}^2},
\ee
and its uniform and fast convergence.

The phase factor $e^{-iS(\tau)}$ in (\ref{92'}) is given by
\be
S(\tau)
&=&
\epsilon
\frac{\hbar^2}{4\mu}
\left({\cal I}(\sqrt[4]{\mathtt{x}^4-\ell^4\tau},\omega,\tau)+{\cal I}(\sqrt[4]{\mathtt{x}^4-\ell^4\tau},0,\tau)\right).
\ee
Its late-$\tau$ asymptotics  should be compared with $\omega\mathtt{t}/2=\omega\epsilon\hbar\tau/2$ occurring in (\ref{73}).

To this end, we have to recall the support property of the initial condition at $\tau=0$, and its consequence
\be
A_0^4<\mathtt{x}^4-\ell^4\tau<B_0^4,
\ee
where $B_0-A_0$ is of the order of several light-minutes, roughly 1 AU (for a justification of the estimate, see \cite{MC2023}). Under these assumptions, we are interested in the asymptotic form of
\be
{\cal I}(\sqrt[4]{\mathtt{x}^4-\ell^4\tau},\omega,\tau)
&=&
-\frac{\hbar}{2\mu\omega}
\frac{1}{\ell^4}
\ln\left(
\sqrt{1+\frac{4\mu^2\omega^2}{\hbar^2}(\mathtt{x}^4-\ell^4\tau)}+\frac{2\mu\omega}{\hbar}\sqrt{\mathtt{x}^4-\ell^4\tau}
\right)
-
\frac{1}{\ell^4}
\sqrt{1+\frac{4\mu^2\omega^2}{\hbar^2}(\mathtt{x}^4-\ell^4\tau)}\sqrt{\mathtt{x}^4-\ell^4\tau}
\nonumber\\
&\pp=&
+
\frac{\hbar}{2\mu\omega}
\frac{1}{\ell^4}
\ln\left(
\sqrt{1+\frac{4\mu^2\omega^2}{\hbar^2}\mathtt{x}^4}+\frac{2\mu\omega}{\hbar}\mathtt{x}^2
\right)
+
\frac{1}{\ell^4}
\sqrt{1+\frac{4\mu^2\omega^2}{\hbar^2}\mathtt{x}^4}\mathtt{x}^2,
\\
{\cal I}(\sqrt[4]{\mathtt{x}^4-\ell^4\tau},0,\tau)
&=&
-\frac{2}{\ell^4}
\left(
\sqrt{\mathtt{x}^4-\ell^4\tau}
-
\mathtt{x}^2
\right),
\ee
which, effectively, can be reduced by means of the gap-hyperboloid condition to $\mathtt{x}\approx \ell\tau^{1/4}$,
\be
{\cal I}(\sqrt[4]{\mathtt{x}^4-\ell^4\tau},\omega,\tau)
+
{\cal I}(\sqrt[4]{\mathtt{x}^4-\ell^4\tau},0,\tau)
&\approx&
\frac{\hbar}{2\mu\omega}
\frac{1}{\ell^4}
\ln\left(
\sqrt{1+\frac{4\mu^2\omega^2}{\hbar^2}\mathtt{x}^4}+\frac{2\mu\omega}{\hbar}\mathtt{x}^2
\right)
\nonumber\\
&\pp=&+
\frac{1}{\ell^4}
\sqrt{1+\frac{4\mu^2\omega^2}{\hbar^2}\mathtt{x}^4}\mathtt{x}^2
+
\frac{2}{\ell^4}
\mathtt{x}^2
\\
&\approx&
\frac{\hbar}{2\mu\omega}
\frac{1}{\ell^4}
\ln\left(
\sqrt{\frac{4\mu^2\omega^2}{\hbar^2}\mathtt{x}^4}+\frac{2\mu\omega}{\hbar}\mathtt{x}^2
\right)
+
\frac{1}{\ell^4}
\sqrt{\frac{4\mu^2\omega^2}{\hbar^2}\mathtt{x}^4}\mathtt{x}^2
+
\frac{2}{\ell^4}
\mathtt{x}^2
\\
&\approx&
\frac{\hbar}{2\mu\omega}
\frac{1}{\ell^4}
\ln\left(
\frac{4\mu\omega}{\hbar}\mathtt{x}^2
\right)
+
\frac{1}{\ell^4}
\frac{2\mu\omega}{\hbar}\mathtt{x}^4
+
\frac{2}{\ell^4}
\mathtt{x}^2
\\
&\approx&
\frac{1}{\ell^4}
\frac{2\mu\omega}{\hbar}\mathtt{x}^4
\approx 
\frac{1}{\ell^4}
\frac{2\mu\omega}{\hbar}\ell^4\tau.
\ee
Asymptotically, for large $\tau$, we obtain the expected result,
\be
S(\tau)
&\approx&
\epsilon
\frac{\hbar^2}{4\mu}\frac{2\mu\omega}{\hbar}\tau
=
\frac{\omega}{2}\epsilon
\hbar\tau
=
\frac{\omega}{2}\mathtt{t}.
\ee

\section{$1+3\to 3$ reduction: three-space probabilities}
\label{Sec6}

The three-space probability density is defined by either 
\be
\rho_\tau(\xi,\theta,\varphi)
&=&
\int
d\mathtt{x}\,\mathtt{x}^3\sinh^2\xi\sin\theta
|\psi_\tau(\mathtt{x},\xi,\theta,\varphi)|^2,
\ee
or
\be
\rho_\tau(\xi,\theta,\varphi)
&=&
\int
d\mathtt{x}\,\mathtt{x}^3\xi^2\sin\theta
|\psi_\tau(\mathtt{x},\xi,\theta,\varphi)|^2,
\ee
if we work in the approximation $\sinh\xi\approx\tanh\xi\approx\xi$ (or in a flat universe).
In virtue of the initial condition we assume,  the support of $\psi_\tau(\mathtt{x},\xi,\theta,\varphi)$ is restricted by 
the inequality
\be
\sqrt[4]{A_0^4+\ell^4\tau} \le \mathtt{x}\le \sqrt[4]{B_0^4+\ell^4\tau},\label{112}
\ee
for some $0\le A_0<B_0<\infty$. When we speak of the support we mean, of course,  the closure of the set of those $x^\mu$ where the wave function is non-zero. Hence, even for $A_0=0$ and $\tau=0$, we may treat the argument $x^\mu$ of the wave function as a future-pointing timelike world vector, with $\mathtt{x}$ strictly positive.

A practical implication of inequality (\ref{112}) is that asymptotically, for large $\tau$, the solution is localized in a future-neighborhood of the hyperboloid $\mathtt{x}^2=\ell^2\sqrt{\tau}$, with $\tau$ counted out since the origin of the universe. This, on the other hand, implies that the present age of the universe, when referred to our human labs, is approximately equal to $\ell\sqrt[4]{\tau}$.

For large $\tau$, the theory reconstructs standard quantum mechanics if we treat $\mathtt{r}=\mathtt{x}\xi$ as the measure of distance in position space. More precisely, $\mathtt{r}$ should be treated as the radial coordinate in spherical coordinates. However, the integration over $\mathtt{x}$ implies that $\mathtt{r}=\mathtt{x}\xi$ will not occur in the asymptotic three-space formulas. Therefore, in order to compare the three-space theory with standard 3D-space quantum mechanics, we have to introduce a parameter, $R$, representing an average $\mathtt{x}$, averaged under the assumption of (\ref{112}). Present-day quantum measurements may be expected to involve $R$ of the order of 10-20 light-years. Accordingly, as another rule of thumb, we may assume that   $r=R\xi\approx \xi \ell\sqrt[4]{\tau}$ is the radial coordinate known from quantum mechanics textbooks. 
At time scales $\delta\tau/\tau\ll 1$, available in our human galaxy-scale quantum measurements, we can assume $\xi \ell\sqrt[4]{\tau+\delta\tau}\approx\xi \ell\sqrt[4]{\tau}$. A variation of $r$ with $\tau$ can be ignored as long as the asymptotic form of quantum mechanics is being used.

What we have just described is the correspondence principle with standard quantum mechanics. It is similar to the one introduced by Infeld and Schield \cite{Infeld1945} in their analysis of the Kepler problem. The difference is that \cite{Infeld1945} treats the hyperbolic space as the configuration space for 3-dimensional  position-representation quantum mechanics, whereas in our formalism the configuration space is Minkowskian (i.e. (1+3)-dimensional), and instead of a single hyperbolic geometry the configuration space is a quantum superposition of different hyperbolic geometries (with different curvatures).

\subsection{Approximate three-space probabilities}

We again begin with  the approximation $\sinh\xi\approx\tanh\xi\approx\xi$. $F_{0,0}(\mathtt{x},0)$ is being given by a function of the form depicted in Fig.~\ref{Fig0}, with large $\alpha$, so that the differences with respect to the characteristic function of $]A_0,B_0[$ can be ignored. 
For $l=0$ the dependence on spherical angles is trivial, so we are left with
\be
\rho_\tau(\xi)
&=&
\int_0^\infty d\mathtt{x}\,\mathtt{x}^3\xi^2\,
|\phi_{0,\tau}(\mathtt{x},\xi)|^2\\
&=&
\frac{3}{(B_0^3-A_0^3)\sqrt{\pi}}\left(\frac{\hbar}{\mu\omega}\right)^{3/2}
\frac{e^{-\frac{\mu  \omega}{\hbar}\xi ^2 A_0^2}   \left(\frac{\mu\omega}{\hbar}\xi ^2A_0^2   +1\right)
-
e^{-\frac{\mu  \omega}{\hbar}\xi ^2 B_0^2}   \left(\frac{\mu\omega}{\hbar}\xi ^2B_0^2   +1\right)}{\xi ^4}.
\label{111}
\ee
In order to switch from the shape variable $\xi$ to the asymptotic spherical coordinate $r=R\xi$ (not to be confused with $\mathtt{r}=\mathtt{x}\xi$), we employ the change of variables
\be
\varrho_0(r) 
&=&
R^{-1}\rho_0(R^{-1}r)
\\
&=&
\frac{3 R^3}{(B_0^3-A_0^3)\sqrt{\pi}}\left(\frac{\hbar}{\mu\omega}\right)^{3/2}
\frac{e^{-\frac{A_0^2}{R^2}\frac{\mu  \omega}{\hbar}r^2}   \left(\frac{A_0^2}{R^2}\frac{\mu\omega}{\hbar}r^2   +1\right)
-
e^{-\frac{B_0^2}{R^2}\frac{\mu  \omega}{\hbar}r^2}   \left(\frac{B_0^2}{R^2}\frac{\mu\omega}{\hbar}r^2   +1\right)}{r^4}.
\label{118}
\ee
To illustrate the form of $\varrho_0(r)$, let us take $A_0=0$ and denote by $\tilde\mu=\mu B_0^2/R^2$  the ``renormalized mass''. The resulting density, 
\be
\varrho_0(r) 
&=&
\frac{3}{\sqrt{\pi}}\left(\frac{\hbar}{\tilde\mu\omega}\right)^{3/2}
\frac{1
-
e^{-\frac{\tilde\mu  \omega}{\hbar}r^2}   \left(\frac{\tilde\mu\omega}{\hbar}r^2   +1\right)}{r^4},
\label{120}
\ee
is plotted in Fig.~\ref{Fig3}, as compared to the Gaussian with the same parameters,
\be
\rho_{\textrm{g}}(r)
&=&
2\sqrt{\frac{\tilde\mu\omega}{\pi\hbar}}e^{-\frac{\tilde\mu\omega}{\hbar}r^2}.
\label{121}
\ee
For a given $\tau$, the universe is localized in a future neighborhood of the hyperboloid $\mathtt{x}=\ell \sqrt[4]{\tau}$, so that for a negligible ratio $|\bm x|/\mathtt{x}$ (typical of our-galaxy labs), the Minkowski-space time coordinate of quantum events,  $x^0=ct$, is approximately equal to $\ell \sqrt[4]{\tau}$, a fact implying that
\be
\tilde\mu &=& \mu \frac{B_0^2}{R^2}
\approx
\mu \frac{B_0^2}{\ell^2\sqrt{\tau}}
\approx
\mu \frac{B_0^2}{c^2t^2}
\approx
\mu \frac{B_0^2}{c^2(t_0+\delta t)^2}
\approx
\mu \frac{B_0^2}{c^2t_0^2}
-
2\mu \frac{B_0^2}{c^2t_0^3}\delta t
=
\mu_0\left(1-\frac{2\delta t}{t_0}\right),
\ee
where $\mu_0$ and $t_0$ denote, respectively, the current value of mass of the oscillator and the current age of the universe. $\delta t$ is the duration of the quantum measurement. Assuming $t_0$ is 10 billion years and $\delta t$ a thousand years, we obtain
\be
\tilde \mu 
\approx
\mu_0-\delta\mu_0,
\ee
with $\delta \mu_0/\mu_0\sim 10^{-7}$. The masses we are dealing with have decreased during the past millenium by some $10^{-5}$ percent.

Of course, one should not treat the above estimate too seriously --- we are still at the level of an approximate toy model, with the universe ``filled'' with a single harmonic oscillator. 
\begin{figure}
\includegraphics[width=8 cm]{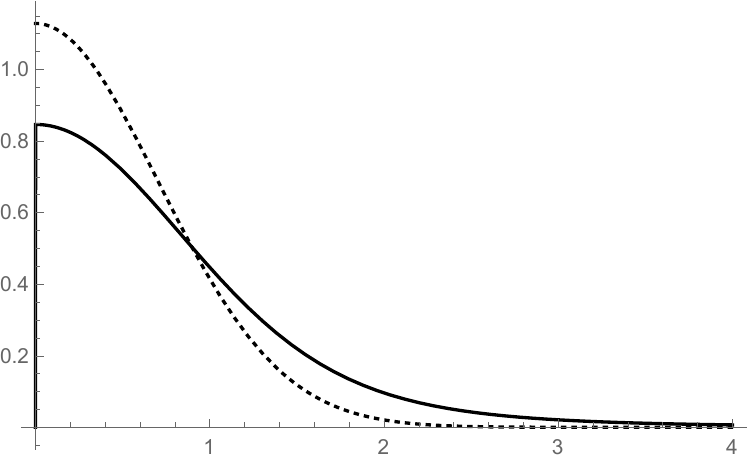}
\caption{Gaussian  (\ref{121}) (dotted) versus $\varrho_0(r)$ (full) given by (\ref{120}). The units are dimensionless, $\tilde\mu\omega/\hbar=1$, and the normalization is $\int_0^\infty dr\,\rho_{\textrm{g}}(r)=\int_0^\infty dr\,\varrho_{0}(r)=1$. The parameter that controls both densities, $\tilde\mu=\mu B_0^2/R^2$, effectively depends on $\tau$, because asymptotically $R\approx \ell\sqrt[4]{\tau}$. The fact that wave functions are defined on space-time makes the reduced ground state additionally smeared out in the 3-position space. The  effective $\tau$-dependence of $\tilde\mu$  can be ignored as long as the duration of quantum measurements, $\delta\tau$, is negligible in comparison to the age $\tau$ of the universe, $\delta\tau/\tau\ll1$.}
\label{Fig3}
\end{figure}

\subsection{The exact three-space probabilities}

Assuming that, within the range of integration, $F_{0,0}(\mathtt{x},0)$ is well approximated by a constant $C$, we find
\be
\rho_\tau(\xi)
&=&
\rho_0(\xi)
=
|C|^2
\int_{A_0}^{B_0} d\mathtt{x}\,\mathtt{x}^3\,
(\cosh\xi)^{1-\sqrt{1+\frac{4\mu^2\omega^2\mathtt{x}^4}{\hbar^2}}}
\\
&=&
\frac{\hbar^2}{2\mu^2\omega^2}\frac{|C|^2}{4}
\frac{\cosh\xi-(1+\ln\cosh^z\xi) \cosh^{1-z}\xi}{(\ln\cosh\xi)^2}
\Bigg|_{z=\sqrt{1+\frac{4\mu^2\omega^2A_0^4}{\hbar^2}}}^{\sqrt{1+\frac{4\mu^2\omega^2B_0^4}{\hbar^2}}}
,\label{119}\\
\rho_\tau(0)
&=&
\frac{|C|^2}{4}(B_0^4-A_0^4)
,
\ee
where $f(z)\big|_{z=a}^b=f(b)-f(a)$. 
In order to compare (\ref{119}) with (\ref{111}), without invoking a cumbersome explicit formula for $|C|^2$, let us take $A_0=0$ and express (\ref{111}) in terms of its value at $\xi=0$, 
\be
\rho_\tau(0)
&=&
\frac{3B_0}{2\sqrt{\pi}}\left(\frac{\hbar}{\mu\omega}\right)^{-1/2}=R\varrho_\tau(0),
\\
\varrho_\tau(r)
&=&
\frac{2\hbar^2}{\tilde\mu^2\omega^2}\varrho_\tau(0)
\frac{1
-
e^{-\frac{\tilde\mu  \omega}{\hbar}r^2 }   \left(\frac{\tilde\mu\omega}{\hbar} r^2   +1\right)}{r ^4},\label{122}
\ee
where $\tilde\mu=\mu B_0^2/R^2$. 
Analogously, setting $A_0=0$ in (\ref{119}), we find
\be
\varrho_\tau(r)
&=&
\frac{2\hbar^2}{\tilde\mu^2\omega^2}\varrho_\tau(0)
\frac{(1+\ln\cosh^z(r/R)) \cosh^{1-z}(r/R)}{4R^4(\ln\cosh(r/R))^2}
\Big|_{z=\sqrt{1+\frac{4\tilde\mu^2\omega^2R^4}{\hbar^2}}}^1.
\label{123}
\ee
Now, one can directly verify that (\ref{122}) is the $R\to\infty$ limit of (\ref{123}). The limit is taken with $\tilde\mu={\rm const}$.

\section{A two-body problem}
\label{2-body}

A two-body problem is always implicitly present in problems involving nontrivial potentials. It is also a prelude to more realistic multi-particle systems. 
We begin with
\be
i\dot\psi_{\tau}(\mathtt{x}_1,\xi_1,\theta_1,\varphi_1,\mathtt{x}_2,\xi_2,\theta_2,\varphi_2)
&=&
-i
\ell^4\frac{\partial}{\partial(\mathtt{x}_1^4)}\psi_{\tau}(\mathtt{x}_1,\xi_1,\theta_1,\varphi_1,\mathtt{x}_2,\xi_2,\theta_2,\varphi_2)
\nonumber\\
&\pp=&
-i
\ell^4\frac{\partial}{\partial(\mathtt{x}_2^4)}\psi_{\tau}(\mathtt{x}_1,\xi_1,\theta_1,\varphi_1,\mathtt{x}_2,\xi_2,\theta_2,\varphi_2)
\nonumber\\
&\pp=&
+
\epsilon
\left(
-\frac{\hbar^2}{2m_1}
\Delta_{\mathtt{x}_1}
-\frac{\hbar^2}{2m_2}
\Delta_{\mathtt{x}_2}
+
U(\mathtt{x}_1,\xi_1,\mathtt{x}_2,\xi_2)
\right)
\psi_{\tau}(\mathtt{x}_1,\xi_1,\theta_1,\varphi_1,\mathtt{x}_2,\xi_2,\theta_2,\varphi_2).
\nonumber\\
\ee
Restricting our analysis to the $l=0$ cases, we can separate the angular variables and concentrate on
\be
i\dot\phi_{\tau}(\mathtt{x}_1,\xi_1,\mathtt{x}_2,\xi_2)
&=&
-i
\ell^4\frac{\partial}{\partial(\mathtt{x}_1^4)}\phi_{\tau}(\mathtt{x}_1,\xi_1,\mathtt{x}_2,\xi_2)
-i
\ell^4\frac{\partial}{\partial(\mathtt{x}_2^4)}\phi_{\tau}(\mathtt{x}_1,\xi_1,\mathtt{x}_2,\xi_2)
\nonumber\\
&\pp=&
+
\epsilon
\left(
-\frac{\hbar^2}{2m_1}
\Delta_{\mathtt{x}_1}
-\frac{\hbar^2}{2m_2}
\Delta_{\mathtt{x}_2}
+
U(\mathtt{x}_1,\xi_1,\mathtt{x}_2,\xi_2)
\right)
\phi_{\tau}(\mathtt{x}_1,\xi_1,\mathtt{x}_2,\xi_2),
\ee
where
\be
\Delta_{\mathtt{x}_k}
&=&
\frac{1}{\mathtt{x}_k^2\sinh^2\xi_k}\frac{\partial}{\partial\xi_k}\sinh^2\xi_k\frac{\partial}{\partial\xi_k},
\quad k=1,2.
\ee
Switching to the interaction picture,
\be
\Phi_{\tau}(\mathtt{x}_1,\xi_1,\mathtt{x}_2,\xi_2)
&=&
e^{\tau\ell^4\big(
\frac{\partial}{\partial(\mathtt{x}_1^4)}
+
\frac{\partial}{\partial(\mathtt{x}_2^4)}
\big)}
\phi_{\tau}(\mathtt{x}_1,\xi_1,\mathtt{x}_2,\xi_2)
\\
&=&
\phi_{\tau}\big(\mathtt{x}_1(\tau),\xi_1,\mathtt{x}_2(\tau),\xi_2\big),\\
\mathtt{x}_k(\tau)
&=&
\sqrt[4]{\mathtt{x}_k^4+\ell^4\tau},
\ee
we find
\be
i\dot\Phi_{\tau}(\mathtt{x}_1,\xi_1,\mathtt{x}_2,\xi_2)
&=&
\epsilon
\left(
-\frac{\hbar^2}{2m_1}
\Delta_{\mathtt{x}_1(\tau)}
-\frac{\hbar^2}{2m_2}
\Delta_{\mathtt{x}_2(\tau)}
+
U\big(\mathtt{x}_1(\tau),\xi_1,\mathtt{x}_2(\tau),\xi_2\big)
\right)
\Phi_{\tau}(\mathtt{x}_1,\xi_1,\mathtt{x}_2,\xi_2).
\ee
Now, let
\be
F_{\tau}(\mathtt{x}_1,\xi_1,\mathtt{x}_2,\xi_2)
=
\Phi_{\tau}(\mathtt{x}_1,\xi_1,\mathtt{x}_2,\xi_2)\sinh\xi_1\sinh\xi_2.
\ee
Employing $g^{-2}\big(g^2 (f/g)'\big)'=g^{-1}\big(f''-f g''/g\big)$,
we obtain
\be
i\dot F_{\tau}(\mathtt{x}_1,\xi_1,\mathtt{x}_2,\xi_2)
&=&
\epsilon
\left[
-\frac{\hbar^2}{2m_1\mathtt{x}_1(\tau)^2}
\left(\frac{\partial^2}{\partial\xi_1^2}-1\right)
-\frac{\hbar^2}{2m_2\mathtt{x}_2(\tau)^2}
\left(\frac{\partial^2}{\partial\xi_2^2}-1\right)
+
U\big(\mathtt{x}_1(\tau),\xi_1,\mathtt{x}_2(\tau),\xi_2\big)
\right]
F_{\tau}(\mathtt{x}_1,\xi_1,\mathtt{x}_2,\xi_2).
\nonumber\\
\ee
An analogue of the center-of-mass system of coordinates is defined by
\be
\Xi
&=&
\frac{m_1\mathtt{x}_1(\tau)^2\xi_1+m_2\mathtt{x}_2(\tau)^2\xi_2}
{m_1\mathtt{x}_1(\tau)^2+m_2\mathtt{x}_2(\tau)^2},\\
\xi
&=&
\xi_2-\xi_1.
\ee
Repeating standard calculations, we arrive at the interaction-picture Schr\"odinger equation,
\be
i\dot G_{\tau}(\mathtt{x}_1,\Xi,\mathtt{x}_2,\xi)
&=&
\epsilon
\left(
-\frac{\hbar^2}{2I\big(\mathtt{x}_1(\tau),\mathtt{x}_2(\tau)\big)}
\frac{\partial^2}{\partial\Xi^2}
-\frac{\hbar^2}{2\iota\big(\mathtt{x}_1(\tau),\mathtt{x}_2(\tau)\big)}
\frac{\partial^2}{\partial\xi^2}
+
V\big(\mathtt{x}_1(\tau),\Xi,\mathtt{x}_2(\tau),\xi\big)
\right)
G_{\tau}(\mathtt{x}_1,\Xi,\mathtt{x}_2,\xi),
\nonumber\\
\ee
involving  reduced and total  moments of inertia  (rather than reduced and total masses),
\be
\frac{1}{\iota\big(\mathtt{x}_1,\mathtt{x}_2\big)}
&=&
\frac{1}{m_1\mathtt{x}_1^2}
+
\frac{1}{m_2\mathtt{x}_2^2}
,\\
I\big(\mathtt{x}_1,\mathtt{x}_2\big)
&=&
m_1\mathtt{x}_1^2+m_2\mathtt{x}_2^2,
\ee
the wave function,
\be
G_{\tau}(\mathtt{x}_1,\Xi,\mathtt{x}_2,\xi)
&=&
F_{\tau}\left(\mathtt{x}_1,\Xi -\frac{m_2\mathtt{x}_2^2}{m_2\mathtt{x}_2^2
+
m_1\mathtt{x}_1^2}\xi,\mathtt{x}_2,\Xi +\frac{m_1\mathtt{x}_1^2}{m_2\mathtt{x}_2^2
+
m_1\mathtt{x}_1^2}\xi\right),
\ee
and the potential
\be
V(\mathtt{x}_1,\Xi,\mathtt{x}_2,\xi)
&=&
\frac{\hbar^2}{2\iota(\mathtt{x}_1,\mathtt{x}_2)}
+
U\left(\mathtt{x}_1,\Xi -\frac{m_2\mathtt{x}_2^2}{m_2\mathtt{x}_2^2
+
m_1\mathtt{x}_1^2}\xi,\mathtt{x}_2,\Xi +\frac{m_1\mathtt{x}_1^2}{m_2\mathtt{x}_2^2
+
m_1\mathtt{x}_1^2}\xi\right).
\ee
In the limiting case $m_1\to\infty$, one reconstructs the formalism we have used so far, with
\be
\xi_1
&=&
\Xi,\\
\xi_2
&=&
\Xi +\xi,
\ee
and
\be
i\dot G_{\tau}(\mathtt{x}_1,\Xi,\mathtt{x}_2,\Xi+\xi)
&=&
\epsilon
\left(
-\frac{\hbar^2}{2m_2\mathtt{x}_2(\tau)^2}
\frac{\partial^2}{\partial\xi^2}
+
V\big(\mathtt{x}_1(\tau),\Xi,\mathtt{x}_2(\tau),\Xi+\xi\big)
\right)
G_{\tau}(\mathtt{x}_1,\Xi,\mathtt{x}_2,\Xi+\xi).
\ee
The harmonic oscillator example corresponds to  $\Xi=0$, $m_2=\mu$, $\mathtt{x}_2=\mathtt{x}$, and
\be
V(\mathtt{x}_1,0,\mathtt{x},\xi)=
\frac{\mu \omega^2\mathtt{x}^2\tanh^2\xi}{2}.
\ee
For large values of $\tau$, both $\mathtt{x}_1$ and $\mathtt{x}_2$ are localized in a neighborhood of $\mathtt{x}=\ell\sqrt[4]{\tau}$.

\section{Conclusions}
\label{Conclusions}

The discussed formalism is meant as a unification and generalization of both standard cosmology and quantum mechanics. As opposed to classical cosmology, the universe is not represented by a spatial section of some space-time, but by the support of a wave function propagating through space-time. Quantum mechanics known from textbooks is reconstructed asymptotically, for large times, by means of an appropriate correspondence principle. The universe is in general deformed by the presence of matter.  We have decided to perform an explicit analysis of a simple but physically meaningful and exactly solvable system, hence the choice of a harmonic oscillator. Among various possibilities we have chosen the CRS model of quantum harmonic oscillator, very natural in the context of spaces with constant curvature. Yet, as opposed to the original CRS formalism, the curvature in our formalism is not a parameter but a quantum observable. The resulting universe exists in a quantum superposition of different curvatures.

A general conclusion is that, for late times, the evolution of the oscillator is essentially the one we know from standard quantum mechanics, but with one important subtlety. Namely, the effective renormalized mass of the oscillator (inferred on the basis of the uncertainty of its geodesic distance $\mathtt{r}$) is time-dependent, as opposed to the bare mass that occurs in the Hamiltonian. The time in $e^{-i\omega(n+1/2) \mathtt{t}}$  is asymptotically (i.e. for late times) proportional to the quantum evolution parameter, $\mathtt{t}\sim\tau$, the age of the universe is proportional to $\sqrt[4]{\tau}$, and the renormalized mass decays as $1/\sqrt{\tau}$. The effect does not seem to be a peculiarity of this concrete potential. Rather, it is a consequence of the concrete form of the empty-universe Hamiltonian and its coupling with matter.  Since the renormalization of mass is influenced by the dynamics of the universe, the effect may be regarded as yet another version of Mach's principle.

\acknowledgments

Calculations were carried out at the Academic Computer Center in Gda{\'n}sk. The work was supported by the CI TASK grant `Non-Newtonian calculus with interdisciplinary applications'.

\section*{Appendices }

\subsection{An alternative definition of the harmonic oscillator}

In the symmetry scattering approach to Schr\"odinger equations \cite{Wehrhahn}, a potential is identified with the angular part of an appropriate Laplace-Beltrami operator, hence, in our case, this would be the term proportional to $l(l+1)$. Of course, this is not what we are interested in if we want to define a harmonic oscillator. Let us therefore try an alternative, but equally general formulation. 

We start with the observation that in nonrelativistic quantum mechanics we find
\be
\Delta U(\bm x)
&=&
\Delta \frac{m\omega^2 \bm x^2}{2}=3m\omega^2=\textrm{const}>0.
\ee
An analogous generalization can be formulated for any Laplace-Beltrami operator, in particular
\be
\Delta_\mathtt{x}U(\mathtt{x},\xi,\theta,\varphi)
&=&
\frac{1}{\mathtt{x}^2\sinh^2\xi}\frac{\partial}{\partial\xi}\sinh^2\xi\frac{\partial}{\partial\xi}
U(\mathtt{x},\xi,\theta,\varphi)
+
\frac{1}{\mathtt{x}^2\sinh^2\xi}
\left(
\frac{1}{\sin\theta}\frac{\partial}{\partial\theta}\sin\theta\frac{\partial}{\partial\theta}
+
\frac{1}{\sin^2\theta}\frac{\partial^2}{\partial\varphi^2}
\right)U(\mathtt{x},\xi,\theta,\varphi)
\nonumber
\\
&=&
\textrm{const}>0.
\ee
So, first of all,
\be
U(\mathtt{x},\xi,\theta,\varphi)=\mathtt{x}^2V(\xi,\theta,\varphi),
\ee
and
\be
\frac{1}{\sinh^2\xi}\frac{\partial}{\partial\xi}\sinh^2\xi\frac{\partial}{\partial\xi}
V(\xi,\theta,\varphi)
+
\frac{1}{\sinh^2\xi}
\left(
\frac{1}{\sin\theta}\frac{\partial}{\partial\theta}\sin\theta\frac{\partial}{\partial\theta}
+
\frac{1}{\sin^2\theta}\frac{\partial^2}{\partial\varphi^2}
\right)V(\xi,\theta,\varphi)
=
C>0
\ee
%UUUUUUUUUUUUUUUUUUUUUUUUUUUUU
Assuming rotational invariance we get
\be
\frac{1}{\sinh^2\xi}\frac{\partial}{\partial\xi}\sinh^2\xi\frac{\partial}{\partial\xi}
V(\xi)
=
C,
\ee
whose general solution reads
\be
V(\xi)
&=&
\frac{C}{2}\xi\coth\xi+C_1 \coth\xi+C_2.
\ee
For $C=0$ we reconstruct the Infeld-Schield version of the Coulomb-Newton potential \cite{Infeld1945}. It is intriguing that the oscillator and the Kepler problem are so intimately related, even at such a general level.

A finite value at $\xi= 0$ implies $C_1=0$, and then
\be
\lim_{\xi\to 0}
V(\xi)
&=&
\frac{C}{2}+C_2.
\ee
We assume $C_2=-C/2$, so that $V(0)=0$. Then,
\be
U(\mathtt{x},\xi,\theta,\phi)
&=&
\frac{C}{2}\mathtt{x}^2(\xi\coth\xi-1)
=
\frac{C}{2}\frac{\mathtt{x}^2\xi^2}{3}+\dots
\ee
The correspondence principle with standard quantum mechanics implies
\be
H
&=&
-\frac{\hbar^2}{2m}\Delta_\mathtt{x}+\frac{3}{2}m\omega^2\mathtt{x}^2(\xi\coth\xi-1)
\label{156}
\\
&=&
-\frac{\hbar^2}{2m}\Delta_\mathtt{x}+\frac{m\omega^2\mathtt{r}^2}{2}+\dots
\ee
where $\mathtt{r}=\mathtt{x}\xi$ is the geodesic coordinate, i.e.
\be
\Delta_\mathtt{x}U(\mathtt{x},\xi,\theta,\varphi)
&=&
3m\omega^2,
\ee
which is the same definition as in nonrelativistic quantum mechanics. For large $\xi$ the potential is linear,
\be
U(\mathtt{x},\xi,\theta,\varphi)
=
\frac{3}{2}m\omega^2\mathtt{x}^2(\xi\coth\xi-1)
\approx
\frac{3}{2}m\omega^2\mathtt{x}^2(\xi-1)
=
\frac{3}{2}m\omega^2\mathtt{x}(\mathtt{r}-\mathtt{x}).
\ee
As opposed to the CRS potential, (\ref{156}) is expected to have  infinitely many bound states and purely discrete spectrum  (cf. \cite{Robinett}).
Unfortunately, I have not managed to factorize (\ref{156}).

\subsection{Excited states for $\sinh\xi\approx\tanh\xi\approx\xi$}

So far, we have concentrated on the ground states, but it is a simple exercise to derive all the excited states as long as we work in the approximate model, valid for $\sinh\xi\approx\tanh\xi\approx\xi$. All the excited states should lead to coherent states and a semiclassical limit of our theory, the problem worthy of a separate study. The exact CRS model is much more complicated, but still exactly solvable by a combination of the results from  \cite{Carinena2011,Carinena2012} with what we describe in the present paper. It must be kept in mind that potentials proportional to $\tanh^2\xi$ are bounded from above and thus only a finite number of ground states is expected to occur. The same subtlety is found  in the Kepler problem, as discussed in \cite{Infeld1945}. 

For simplicity, let us consider $D=1+1$ so that the angular coordinates are absent from the very outset. The main difference between $D=1+1$ and $D=1+3$, $l=0$, is in the form of the free Hamiltonian, a generator of translations of $\mathtt{x}^D$. In case of a Gaussian, the variable $\mathtt{x}^2$ is replaced in interaction picture by $\mathtt{x}^2+\ell^2\tau$, for $D=1+1$, and by  $\sqrt{\mathtt{x}^4+\ell^4\tau}$, for $D=1+3$. In various other respects the two cases are qualitatively similar.

Assume 
\be
U(\mathtt{x}_1,\mathtt{x},\xi)
&=&
\frac{\iota(\mathtt{x})\omega^2 \xi^2}{2},
\ee
where $\iota(\mathtt{x})$ is a moment of inertia to be specified later. 
Switching to the interaction picture,
\be
\tilde\phi_\tau(\mathtt{x}_1,\xi_1,\mathtt{x},\xi)
&=&
e^{\tau\ell^2
\left(
\frac{\partial}{\partial(\mathtt{x}_1^2)}
+
\frac{\partial}{\partial(\mathtt{x}^2)}
\right)
}
\phi_\tau(\mathtt{x}_1,\xi_1,\mathtt{x},\xi)
\\
&=&
\phi_\tau\left(\sqrt{\mathtt{x}_1^2+\tau\ell^2},\xi_1,\sqrt{\mathtt{x}^2+\tau\ell^2},\xi\right),
\ee
we obtain a $\tau$-dependent Hamiltonian,
\be
i\frac{d}{d\tau}\tilde\phi_\tau(\mathtt{x}_1,\xi_1,\mathtt{x},\xi)
&=&
\epsilon
\left(
-\frac{\hbar^2}{2\iota\left(\sqrt{\mathtt{x}^2+\tau\ell^2}\right)}
\frac{\partial^2}{\partial \xi^2}
+
\frac{\iota\left(\sqrt{\mathtt{x}^2+\tau\ell^2}\right)\omega^2 \xi^2}{2}
\right)
\tilde\phi_\tau(\mathtt{x}_1,\xi_1,\mathtt{x},\xi)
\label{36}
\\
&=&
\epsilon
\hbar\omega
\left(
a\left(\sqrt{\mathtt{x}^2+\tau\ell^2}\right)^\dag 
a\left(\sqrt{\mathtt{x}^2+\tau\ell^2}\right)
+
\frac{1}{2}
\right)
\tilde\phi_\tau(\mathtt{x}_1,\xi_1,\mathtt{x},\xi),
\label{37}
\ee
where
\be
a(\mathtt{x})
&=&
\frac{1}{\sqrt{\hbar\omega}}
\left(
\frac{\hbar}{\sqrt{2\iota(\mathtt{x})}}
\frac{\partial}{\partial \xi}
+
\sqrt{\frac{\iota(\mathtt{x})}{2}}\omega\xi
\right),\label{38}\\
a(\mathtt{x})^\dag
&=&
\frac{1}{\sqrt{\hbar\omega}}
\left(
-\frac{\hbar}{\sqrt{2\iota(\mathtt{x})}}
\frac{\partial}{\partial \xi}
+
\sqrt{\frac{\iota(\mathtt{x})}{2}}\omega\xi
\right).\label{39}
\ee
The ground state 
\be
\tilde\phi_{0,\tau}(\mathtt{x}_1,\xi_1,\mathtt{x},\xi)
&=&
\phi_{0,\tau}\left(\sqrt{\mathtt{x}_1^2+\tau\ell^2},\xi_1,\sqrt{\mathtt{x}^2+\tau\ell^2},\xi\right),
\ee
is defined by
\be
a\left(\sqrt{\mathtt{x}^2+\tau\ell^2}\right)
\tilde\phi_{0,\tau}(\mathtt{x}_1,\xi_1,\mathtt{x},\xi)
&=&
a\left(\sqrt{\mathtt{x}^2+\tau\ell^2}\right)
\phi_{0,\tau}\left(\sqrt{\mathtt{x}_1^2+\tau\ell^2},\xi_1,\sqrt{\mathtt{x}^2+\tau\ell^2},\xi\right)
=0.\label{41}
\ee
Following the standard steps one proves
\be
i\frac{d}{d\tau}\tilde\phi_{n,\tau}(\mathtt{x}_1,\xi_1,\mathtt{x},\xi)
&=&
\epsilon
\hbar\omega
\left(n+\frac{1}{2}\right)
\tilde\phi_{n,\tau}(\mathtt{x}_1,\xi_1,\mathtt{x},\xi),
\label{42}\\
\tilde\phi_{n,\tau}(\mathtt{x}_1,\xi_1,\mathtt{x},\xi)
&=&
e^{-i\epsilon
\hbar\omega \left(n+\frac{1}{2}\right)
\tau}
\tilde\phi_{n,0}(\mathtt{x}_1,\xi_1,\mathtt{x},\xi),
\ee
or equivalently,
\be
\phi_{n,\tau}\left(\sqrt{\mathtt{x}_1^2+\tau\ell^2},\xi_1,\sqrt{\mathtt{x}^2+\tau\ell^2},\xi\right)
=
e^{-i\epsilon
\hbar\omega\left(n+\frac{1}{2}\right)
\tau}
\phi_{n,0}(\mathtt{x}_1,\xi_1,\mathtt{x},\xi)
\ee
and thus
\be
\phi_{n,\tau}(\mathtt{x}_1,\xi_1,\mathtt{x},\xi)
=
e^{-i\epsilon
\hbar\omega\left(n+\frac{1}{2}\right)
\tau}
\phi_{n,0}\left(\sqrt{\mathtt{x}_1^2-\tau\ell^2},\xi_1,\sqrt{\mathtt{x}^2-\tau\ell^2},\xi\right),\label{n-phi}
\ee
with
\be
\phi_{n,0}\left(\mathtt{x}_1,\xi_1,\mathtt{x},\xi\right)
&=&
\frac{1}{\sqrt{2^n n!}}
f\left(\mathtt{x}_1,\xi_1,\mathtt{x}\right)H_n\left(\sqrt{\frac{\iota(\mathtt{x})\omega }{\hbar}}\xi\right)
e^{-\frac{\iota(\mathtt{x})\omega}{2\hbar}\xi^2},\\
H_n(\eta)
&=&
(-1)^n
e^{\eta^2}
\frac{d^n}{d\eta^n}
e^{-\eta^2},
\label{46}
\ee
and $f\left(\mathtt{x}_1,\xi_1,\mathtt{x}\right)$ still unspecified. Formula (\ref{n-phi}) shows that the usual quantum mechanical time paramater can be identified with $\mathtt t=\epsilon\hbar\tau$. Note that $\mathtt t$ is as invariant as $\tau$ and $\mathtt x$, and thus cannot be regarded as a timelike coordinate $t=x^0/c$ of a world-position.

Let us recall that in  $m_1\to\infty$ limit the first two coordinates, 
\be
x_1 &=& (x_1^0,x_1^1)
=
\mathtt{x}_1(\cosh\xi_1,\sinh\xi_1), 
\ee
describe the center-of-mass spacetime-position operator of the system. We can consider an initial condition that separates the center of mass coordinate $x_1^\mu$ from $x^\mu$, the relative one,
and concentrate on the latter,
\be
\phi_{n,\tau}(\mathtt{x},\xi)
&=&
e^{-i\epsilon
\hbar\omega\left(n+\frac{1}{2}\right)
\tau}
\frac{1}{\sqrt{2^n n!}}
f\left(\sqrt{\mathtt{x}^2-\tau\ell^2}\right)
H_n\left(\sqrt{\frac{\omega \iota\left(\sqrt{\mathtt{x}^2-\tau\ell^2}\right)}{\hbar}}\xi\right)
e^{-\frac{\omega}{2\hbar}\iota\left(\sqrt{\mathtt{x}^2-\tau\ell^2}\right)\xi^2}.
\label{52'''}
\ee
It is subject to normalization
\be
1 
&=&
\langle \phi_{n,\tau}|\phi_{n,\tau}\rangle
=
\langle \phi_{n,0}|\phi_{n,0}\rangle
=
\langle \phi_{0,0}|\phi_{0,0}\rangle
\\
&=&
\int_{0}^{\infty}d\mathtt{x}\,\mathtt{x} |f(\mathtt{x})|^2
\int_{-\infty}^\infty d\xi
e^{-\frac{\omega}{\hbar}\iota(\mathtt{x})\xi^2}
\label{170'''}\\
&=&
\sqrt{\frac{\pi\hbar}{\omega}}
\int_{0}^{\infty}d\mathtt{x}
\frac{\mathtt{x}|f(\mathtt{x})|^2}{\sqrt{\iota(\mathtt{x})}}.
\label{170''''}\ee
In (\ref{170'''}) we integrate over  $\xi\in\mathbb{R}$ and not over $\xi\in\mathbb{R}_+$, a peculiarity of the 1+1 case.

\subsubsection{Constant moment of inertia, $\iota(\mathtt{x})=m R^2$}

For $\iota(\mathtt{x})=m R^2$ (i.e. in a flat universe), we obtain
\be
\phi_{n,\tau}(\mathtt{x},\xi)
&=&
e^{-i\epsilon
\hbar\omega\left(n+\frac{1}{2}\right)
\tau}
\frac{1}{\sqrt{2^n n!}}
f\left(\sqrt{\mathtt{x}^2-\tau\ell^2}\right)H_n\left(\sqrt{\frac{m\omega }{\hbar}}R\xi\right)
e^{-\frac{m\omega}{2\hbar}R^2\xi^2},\label{50}
\ee
with
\be
A_0^2<\mathtt{x}^2-\ell^2\tau<B_0^2.
\ee
This is, essentially, an oscillator described in terms of some geodesic coordinate $r=R\xi$ and time $\mathtt t=\epsilon\hbar\tau$.

\subsubsection{$\mathtt{x}$-dependent  moment of inertia, $\iota(\mathtt{x})=m\mathtt{x}^2$}

For $\iota(\mathtt{x})=m\mathtt{x}^2$ (i.e. in the hyperbolic universe), the non-vanishing part of the solution reads
\be
\phi_{n,\tau}(\mathtt{x},\xi)
&=&
e^{-i\epsilon
\hbar\omega\left(n+\frac{1}{2}\right)
\tau}
\frac{1}{\sqrt{2^n n!}}
f\left(\sqrt{\mathtt{x}^2-\ell^2\tau}\right)
H_n\left(\sqrt{\frac{m\omega }{\hbar}}\sqrt{\mathtt{x}^2-\ell^2\tau}\xi\right)
e^{-\frac{m\omega}{2\hbar}(\mathtt{x}^2-\ell^2\tau)\xi^2},\label{51}
\ee
with
\be
A_0^2<\mathtt{x}^2-\ell^2\tau<B_0^2.
\ee

\subsubsection{Renormalization of mass}

Define $\mathtt{r}=\mathtt{x}\xi$. For 
$
A_0^2<\mathtt{x}^2-\ell^2\tau<B_0^2
$
rewrite (\ref{52'''}) as
\be
\check\phi_{n,\mathtt{t}}(\mathtt{x},\mathtt{r})
&=&
e^{-i\omega\left(n+\frac{1}{2}\right)\mathtt{t}}
\frac{1}{\sqrt{2^n n!}}
f\left(\sqrt{\mathtt{x}^2-\ell^2\tau}\right)
H_n\left(\sqrt{\frac{m\omega }{\hbar}}\gamma(\mathtt{t},\mathtt{x})\mathtt{r}\right)
e^{-\frac{m\omega}{2\hbar}\gamma(\mathtt{t},\mathtt{x})^2\mathtt{r}^2},\\
\gamma(\mathtt{t},\mathtt{x})
&=&
\sqrt{\frac{\iota(\sqrt{\mathtt{x}^2-\ell c\mathtt{t}})}{m\mathtt{x}^2}}
=
\left\{
\begin{array}{cl}
R/\mathtt{x} & \textrm{for $\iota=mR^2$}\\
\sqrt{1-\ell^2\tau/\mathtt{x}^2} & \textrm{for $\iota=m\mathtt{x}^2$}
\end{array}
\right.
,
\ee
where
\be
A_0/\mathtt{x} <\sqrt{1-\ell^2\tau/\mathtt{x}^2} <B_0/\mathtt{x}.
\ee
Otherwise $\check\phi_{n,\mathtt{t}}(\mathtt{x},\mathtt{r})=0$. 

An observer who performs position measurements in terms of $\mathtt{r}$ will conclude that the role of the parameter $\frac{m\omega }{\hbar}$ is played by its rescaled version $\frac{m\omega }{\hbar}\gamma(\mathtt{t},\mathtt{x})^2$. Simultaneously, one does not observe any modification of frequency in the oscillating term $e^{-i\omega\left(n+\frac{1}{2}\right)\mathtt{t}}$. Accordingly, the rescaling of $m\omega/\hbar$ has to be caused by a renormalization of $m/\hbar$, the parameter that controls the classical limit of the theory. It seems most natural to associate the effect with the renormalization of mass. Therefore, defining
\be
m(\mathtt{t},\mathtt{x})
&=&
m
\gamma(\mathtt{t},\mathtt{x})^2,
\ee
we find that the observed mass $m(\mathtt{t},\mathtt{x})$ decays asymptotically, for late times, as $1/\mathtt{x}^2\sim 1/\mathtt{t}$.  For $D=1+3$ the dependence on $\mathtt{t}$ is different, but still  $m(\mathtt{t},\mathtt{x})$ asymptotically tends to zero.
Observers performing quantum measurements in position representation defined by $\mathtt{r}=\mathtt{x}\xi$, may conclude that the present  mass of the oscillator is smaller than its earlier values.

Let us note that the change of variables from $(\mathtt{x},\xi)$ to $(\mathtt{x},\mathtt{r})$ entails a modification of the form of ${\cal H}_0$. Denoting, 
\be
\psi(\mathtt{x},\xi)
&=&
\Psi(\mathtt{x}\cosh\xi,\mathtt{x}\sinh\xi)=\Psi(x^0,x^1),\\
\check \psi(\mathtt{x},\mathtt{r})
&=& 
\Psi\left(\mathtt{x}\cosh\frac{\mathtt{r}}{\mathtt{x}},\mathtt{x}\sinh\frac{\mathtt{r}}{\mathtt{x}}\right),
\ee
and employing the proportionality between ${\cal H}_0$ and Euler's homogeneity operator,
we find
\be
{\cal H}_0\psi(\mathtt{x},\xi) &=&
-i
\ell^2\frac{\partial}{\partial(\mathtt{x}^2)}\psi(\mathtt{x},\xi),\label{182'''}
\\
{\cal H}_0\check \psi(\mathtt{x},\mathtt{r})
&=&
-i
\frac{\ell^2}{2\mathtt x^2}
\left(
\mathtt{x}\frac{\partial}{\partial\mathtt{x}}
+
 \mathtt{r}\frac{\partial}{\partial \mathtt{r}}
\right)\check \psi(\mathtt{x},\mathtt{r}).\label{183'''}
\ee
The inverse formulas read
\be
\mathtt{x}
&=&
\sqrt{(x^0)^2-(x^1)^2},\\
\xi&=&
\ln\sqrt{\frac{x^0+x^1}{x^0-x^1}},
\\
\mathtt{r}&=&
\sqrt{(x^0)^2-(x^1)^2}\ln\sqrt{\frac{x^0+x^1}{x^0-x^1}},\label{r(x)}
\\
\Psi(x^0,x^1)
&=&
\psi\left(\sqrt{(x^0)^2-(x^1)^2},\ln\sqrt{\frac{x^0+x^1}{x^0-x^1}}\right),
\ee
whereas the corresponding  scalar products are related by
\be
\langle \Psi|\Phi\rangle
&=&
\int_{V_+}dx\, \overline{\Psi(x)}\Phi(x)
\\
&=&
\int_0^\infty d\mathtt{x} \int_{-\infty}^\infty d\mathtt{r} \,
\overline{\check\psi(\mathtt{x},\mathtt{r})}\check\phi(\mathtt{x},\mathtt{r})
\\
&=&
\int_0^\infty d\mathtt{x}\,\mathtt{x} \int_{-\infty}^\infty d\xi \,
\overline{\psi(\mathtt{x},\xi)}\phi(\mathtt{x},\xi).
\ee
\begin{figure}
\includegraphics[width=8 cm]{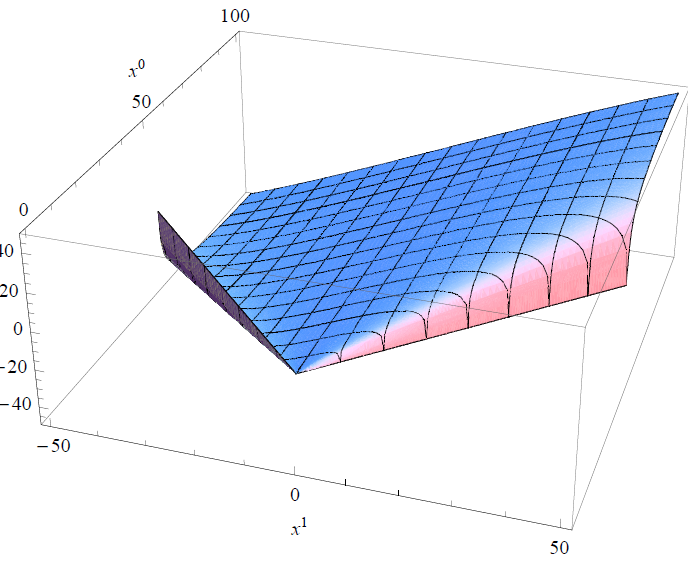}
\caption{Geodesic coordinate $\mathtt{r}(x^0,x^1)$ given by (\ref{r(x)}) as a function of the Minkowski-space coordinates. $\mathtt{r}(x^0,x^1)$ decays to 0 as $x$ approaches the light cone $\mathtt{x}=0$, a property that does not contradict the normalization by means of $\int_{-\infty}^\infty d\mathtt{r}$, because the  integral is computed along the hyperboloid $\mathtt{x}=\textrm{const}>0$. Non-relativistic, late-time asymptotics of the position operator follows from the pointwise but non-uniform convergence $\lim_{x^0\to\infty}\mathtt{r}(x^0,x^1)=x^1$. The boundary condition, $\Psi_\tau(x)=0$ if $\mathtt{x}=0$, implies that the light cone does not belong to the universe.}
\label{Fig3}
\end{figure}

\subsubsection{Example: mass vs. effective renormalized mass (hyperbolic case)}

Consider  (\ref{52'''}), with $n=0$, $\iota=m\mathtt{x}^2$ (the hyperbolic case), $A_0=0$, $B_0>0$, 
\be
\phi_{0,\tau}(\mathtt{x},\xi)
&=&
e^{-\frac{i}{2}\epsilon
\hbar\omega
\tau}
f\big(\sqrt{\mathtt{x}^2-\tau\ell^2}\big)
e^{-\frac{\omega m}{2\hbar}(\mathtt{x}^2-\tau\ell^2)\xi^2}
\\
&\approx&
Ce^{-\frac{i}{2}\omega\mathtt{t}}
e^{-\frac{\omega}{2\hbar}m\big(1-\tau\ell^2/\mathtt{x}^2\big)\mathtt{r}^2}, \quad \textrm{for $0<\mathtt{x}^2-\tau\ell^2<B_0^2$}
\label{52''''}
\ee
and 0 otherwise, normalized by (\ref{170''''}),
\be
1 
&=&
\sqrt{\frac{\pi\hbar}{m\omega}}
\int_{0}^{\infty}d\mathtt{x}
|f(\mathtt{x})|^2
\approx
\sqrt{\frac{\pi\hbar}{m\omega}}B_0 |C|^2.
\ee
The renormalized mass,
\be
\tilde m=m\big(1-\tau\ell^2/\mathtt{x}^2\big),
\ee
is the effective mass as measured by the width of the Gaussian, hence by the uncertainty of the geodesic coordinate $\mathtt{r}=\mathtt{x}\xi$. Now assume that $\tau_0$ is the current age of the universe and thus 
\be
0<\tilde m_0=m\big(1-\tau_0\ell^2/\mathtt{x}^2\big)<B_0^2/\mathtt{x}^2,
\ee
is the current renormalized mass of, say, an electron oscillating with frequency $\omega$. Note that $\mathtt{x}$ belongs to the support of the wave function at the current value of $\tau_0$. Denoting $\mathtt{t}_0=\epsilon\hbar\tau_0$,
$\mathtt{t}_0+\delta\mathtt{t}=\epsilon\hbar\tau_0+\epsilon\hbar\delta\tau$, we find
\be
\phi_{0,\tau_0+\delta\tau}(\mathtt{x},\xi)
&\approx&
Ce^{-\frac{i}{2}\omega(\mathtt{t}_0+\delta\mathtt{t})}
e^{-\frac{\omega}{2\hbar}m\big(1-(\tau_0+\delta\tau)\ell^2/\mathtt{x}^2\big)\mathtt{r}^2}
\\
&\approx&
Ce^{-\frac{i}{2}\omega(\mathtt{t}_0+\delta\mathtt{t})}
e^{-\frac{\omega}{2\hbar}\tilde m_0\mathtt{r}^2},
\label{196''}
\ee
if $\delta\mathtt{t}/\mathtt{t}_0\ll 1$, which can be safely assumed for all quantum mechanical experiments performed during the past century. (\ref{196''}) is the standard nonrelativistic result. But if $\delta\tau\approx -\tau_0$, that is we look at the state of the oscillator in a distant past, hence for small $\mathtt{x}$, the renormalized mass tends towards the bare mass, $\tilde m\approx m$, which is much larger. 

\begin{figure}
\includegraphics[width=8 cm]{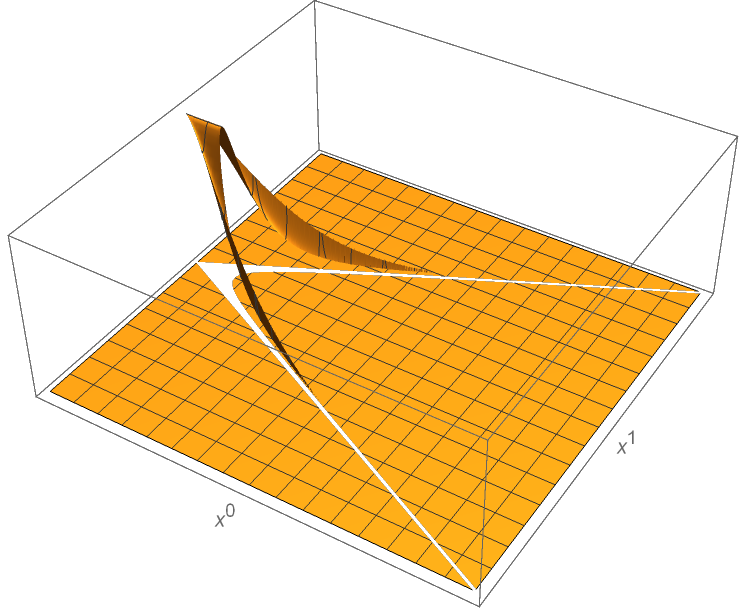}
\includegraphics[width=8 cm]{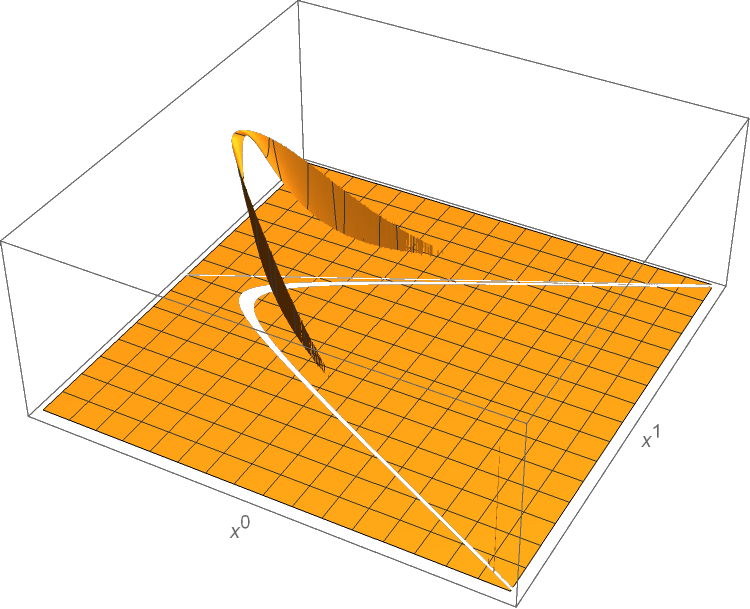}
\caption{An analogue of Fig.~\ref{Fig1} but for the solution (\ref{198''}). The parameters are $B_0=1$, $m\omega/\hbar=1$, $\ell=1$, $\tau=0$ (left), and $\tau=2$ (right). The white region represents the support of the wave function, i.e. the configuration-space  universe as a subset of the Minkowski space in $D=1+1$, characterized by $\Phi_{0,\tau}(x^0,x^1)\neq 0$. The mass is given by its bare value. }
\label{Fig4}
\end{figure}

The situation changes if we work with the bare mass. When expressed in space-time coordinates, $(x^0,x^1)$, the solution evolves in configuration space-time as a squeezed state:
\be
\Phi_{0,\tau}(x^0,x^1)
&=&
e^{-\frac{i}{2}\omega\mathtt{t}}
f\big(\sqrt{(x^0)^2-(x^1)^2-\tau\ell^2}\big)
e^{-\frac{\omega m}{2\hbar}\big((x^0)^2-(x^1)^2-\tau\ell^2\big)\ln^2\sqrt{\frac{x^0+x^1}{x^0-x^1}}}
\\
&\approx&
Ce^{-\frac{i}{2}\omega\mathtt{t}}
e^{-\frac{\omega m}{2\hbar}\big((x^0)^2-(x^1)^2-\tau\ell^2\big)\ln^2\sqrt{\frac{x^0+x^1}{x^0-x^1}}},
\label{198''}
\ee
if $0<(x^0)^2-(x^1)^2-\tau\ell^2<B_0^2$, and vanishes otherwise. Fig.~\ref{Fig4} illustrates the probability density associated with (\ref{198''}) for some arbitrarily chosen dimensionless parameters.

\end{document}